\documentclass[pra,showpacs,twocolumn,superscriptaddress,floatfix,aps]{revtex4-2}
\usepackage{amsmath,amssymb,bm}
\usepackage{graphicx}
\usepackage{physics}
\usepackage[T1]{fontenc}
\usepackage{comment}
\usepackage{mathtools}
\usepackage[normalem]{ulem}
\usepackage{hyperref}
\hypersetup{colorlinks,%,%
    linkcolor=blue,%
    citecolor=blue,%
    urlcolor=blue}
\usepackage{color}
\usepackage{tikz}
\usetikzlibrary{shapes}

\newcommand{\markerone}{\raisebox{-0.1pt}{\tikz{\node[draw,scale=0.6,circle,fill=black](){};}}}
\newcommand{\markertwo}{\raisebox{-0.1pt}{\tikz{\node[draw,scale=0.6,regular polygon, regular polygon sides=4,fill=red!100!red,rotate=180](){};}}}
\newcommand{\markerthree}{\raisebox{-0.1pt}{\tikz{\node[draw,scale=0.6,regular polygon, regular polygon sides=6,fill=white!20!blue,rotate=90](){};}}}
\begin{document}

\title{Quench induced chaotic dynamics of Anderson localized interacting Bose-Einstein condensates in one dimension}

\author{Swarup Sarkar}
%\email{skanti@iitg.ac.in}
\affiliation{Department of Physics, Indian Institute of Technology Guwahati, Guwahati 781039, Assam, India}

\author{Tapan Mishra}
%\email{mishratapan@gmail.com}
\affiliation{School of Physical Sciences, National Institute of Science Education and Research, Jatni 752050, India}
\affiliation{Homi Bhabha National Institute, Training School Complex, Anushaktinagar, Mumbai 400094, India}

\author{Paulsamy Muruganandam}
%\email{anand@bdu.ac.in}
\affiliation{Department of Physics, Bharathidasan University, Tiruchirappalli 620024, Tamilnadu, India}

\author{Pankaj K. Mishra}
%\email{pankaj.mishra@iitg.ac.in}
\affiliation{Department of Physics, Indian Institute of Technology Guwahati, Guwahati 781039, Assam, India}
\date{\today}

\begin{abstract} 

We study the effect of atomic interaction on the localization and the associated dynamics of Bose-Einstein condensates in a one-dimensional quasiperiodic optical lattice and random Gaussian disordered potentials. When the interactions are absent, the condensates exhibit localization, which weakens as we increase the interaction strength beyond a threshold value for both potential types. We inspect the localized and delocalized states by perturbing the system via quenching the interaction strength instantaneously to zero and studying the dynamics of the condensate, which we further corroborate using the out-of-time-order correlator. The temporal behaviour of the time correlator displays regular dynamics for the localized state, while it shows temporal chaos for the delocalized state. We confirm this dynamical behaviour by analyzing the power spectral density of the time correlator. We further identify that the condensate admits a quasiperiodic route to chaotic dynamics for both potentials. Finally, we present the variation of the maximal Lyapunov exponents for different nonlinearity and disorder strengths that have a positive value in the regime where the time correlator function shows chaotic behaviour. Through this, we establish the strong connection between the spatially delocalized state of the condensate and its temporal chaos.
\end{abstract}

\maketitle

\section{Introduction}
\label{sec1}

Localization of matter waves in random media has been the topic of interest in condensed matter physics in the last several decades~\cite{Modugno2010, Lagendijk2009,  Aspect2009}. Since the seminal prediction of the exponential localization of the electronic wavefunction in the presence of random disorder by P. W. Anderson (known as Anderson localization~\cite{Anderson1958}), the phenomenon of localization has attracted significant attention. There were several efforts to observe localization in various systems, such as electromagnetic waves~\cite{Wiersma1997, Scheffold1999, Schwartz2007, Aegerter07, Topolancik2007}, microwaves~\cite{Dalichaouch1991, Dembowski1999, Chabanov2000, Pradhan2000}, and acoustic waves~\cite{WEAVER1990,Hu2008, Manley2014}. 
The experimental observations of matter wave localization in 1D~\cite{Moore1994} and 3D~\cite{Chabe2008} kicked rotors have generated significant interest in the field of ultracold matter. This field has shed light on many complex phenomena, including quantum chaos.
%Interestingly experimental observation of matter wave localization in 1D ~\cite{Moore1994} and 3D ~\cite{Chabe2008} kicked rotor seeded some extra interests in the field of ultracold matter which happened to be torchbearer for many complex phenomena like quantum chaos
 On the other hand, in case of non-interacting Bose-Einstein condensates (BECs) of $^{87}$Rb atoms localization was observed after releasing the condensate onto a 1D waveguide created by laser speckle~\cite{Billy2008}. Later on, Roati \textit{et al.} discovered the localization of matter waves in BECs of $^{39}$K atoms trapped in a 1D bichromatic optical lattice~\cite{Roati2008}. Further, White \textit{et al.} reported a similar localization for the 2D non-interacting condensates of $^{87}$Rb atoms trapped in a point-like disordered potential~\cite{White2020}. Skipetrov \textit{et al.} extended the analysis of the localization of condensates to 3D random potentials~\cite{Skipetrov2008}. After the experimental observations, numerous numerical and theoretical studies have been performed in the recent past that show the localization of the matter wave for weak nonlinearity for the condensate trapped in the quasi-periodic potential~\cite{Adhikari2009, Muru2010, Chengsm2010, Cheng:sp-sm2010, Cheng2011, Cheng:per2011, Cheng2014, Li2016,Abrams2008} and random speckle potential~\cite{deissler2010, Cheng2010, cardoso2012, xi2015, Zhang2022}. There are some numerical~\cite{Pikovsky2008} and experimental~\cite{ Lucioni2011} works that show the destruction of localization in presence of the interactions in the condensate, and also report sub-diffusive nature of the delocalized state~\cite{Pikovsky2008,Kopidakis2008,Lucioni2011}. Cherroret \textit{et al.} demonstrated theoretically that even weak interactions among particles can disrupt the transition from the subdiffusive regime to the transport inhibited regime, also known as the Anderson transition, for expanding localized wave packets in 3D disordered potentials~\cite{Cherroret2014}.

On the other hand, considerable attention has been paid in understanding the dynamical behaviour of matter waves out of equilibrium. Several theoretical \cite{Adhikari2009, Cheng2011} and experimental~\cite{Roati2008, Lucioni2011} studies have been performed on the non-equilibrium dynamics of BECs released from an external trap in different scenarios. For instance, Doggen and Kinnunen reported the transition from the localized to the delocalized state by quenching the nonlinearity from finite value to  zero~\cite{Doggen2014}. Efforts have also been made to understand the dynamics of the localized BEC trapped in disordered optical lattices.
 
In this context, a wealth of novel scenarios has been explored in both theoretical and numerical fronts for both interacting and non-interacting condensates trapped in disordered potentials. These scenarios include enhancements in transport properties, dynamical phase transitions from superfluid to Bose glass, and others. Kuhn \textit{et al.} used the perturbative Green function approach to show the significant role played by localization in the diffusion of non-interacting condensates trapped in 2D speckle potential~\cite{Kuhn2005}.  Damski \textit{et al.} demonstrated the dynamical phase transition from a superfluid to Bose-glass for interacting condensates trapped in 2D speckle potential~\cite{Damski2003}. Palencia \textit{et al.} used theoretical and numerical calculations to show that atomic interactions in the condensate play a significant role in the dynamics for a short time scale when the condensate trapped in 1D disordered potential is released from the trap~\cite{sanchez2007}. Further, Lugan \textit{et al.} extended this analysis to the localization of Bogolyubov quasiparticles for interacting BECs~\cite{Lugar2007}. There are some works that report the presence of more complex dynamics of the condensate trapped in the random potential with the spin-orbit coupling. For instance Mardonov \textit{et al.} numerically demonstrated the appearance of the coexistence of two different mechanism of spin-dynamics, namely, spin-precession as well the separation between the spin-component because of anamalous spin-dependent velocities~\cite{Mardonov2015,Mardonov2018,Mardonov2019}.  
%In this context, a wealth of novel scenarios, such as enhancement in the transport properties, dynamical phase transition from the superfluid to Bose glass, etc., have been explored in the theoretical as well as numerical front for both the interacting as well as noninteracting condensates trapped in the disordered potential. Using the perturbative Green function approach,  Kuhn \textit{et al.} showed a significant role played by the localization in the diffusion of the non-interacting condensate trapped in the 2D speckle potential~\cite{Kuhn2005}. Damski \textit{et al.} theoretically demonstrated the dynamical phase transition from superfluid to Bose-glass for the interacting condensate trapped in 2D speckle potential~\cite{Damski2003}. Following this  Palencia \textit{et al.} using the theoretical and numerical calculation showed the atomic interactions in the condensate play quite significant role in the dynamics for short time scale as the condensate trapped in 1D disordered potential released from the trap~\cite{sanchez2007}. Further Lugan \textit{et al.} extended this analysis for the localization of the Bogolyubov Quasiparticles for the interacting BECs~\cite{ Lugar2007}.
%%Change this line 
Interestingly, the dynamical evolution of BECs has also revealed the connection between the chaos resulting from the competing interaction and disorder in the system~\cite{Brezinov2011, Murugan2002, Verma_2012, tosyali2018regular}. Brezinova \textit{et al.}~\cite{Brezinov2011} demonstrated that once BECs trapped under harmonic potential are released from the trap and subject to either periodic or aperiodic (quasi-periodic, random Gaussian disordered potential, etc.) potential, the condensate shows the expansion which exhibits chaotic nature for the potential strength beyond a certain threshold value. In this work, we also aim to demonstrate a similar chaotic nature of the condensate by sudden quenching of the nonlinearity of the condensate to zero. 

The non-equilibrium dynamics is primarily generated in such systems by releasing the BECs from the trap \cite{sanchez2007,Lucioni2011}, by performing a  sudden quench~\cite{Cherroret2021}in either coupling parameters~\cite{ravisankar2020,Gangwar2022}, or nonlinear interactions~\cite{du2022}. At one hand the time of flight techniques have been quite widely used for  distinguishing the nature of the localized and the delocalized condensate. However, in the case of  random potential where the localization state can be more complex phases like, the Bose-glass phase, the Lifshitz phase, etc., distinguishing the localized and the delocalized phases are more challenging and to the best of our knowledge has not been done comprehensively.  In this paper, we use a slightly different technique, which was not followed in earlier studies to analyze the localized and delocalized state of the matter waves in presence of the quasi-periodic as well as the random potential. We implement the complete cessation of the nonlinear interactions of the condensate once the ground state is obtained. This process ensures the temporal dynamics in the condensate, which has been systematically captured by examining the time evolution of the time correlator function, defined as the spatial average of the projection of the wave function at a particular instant on the stationary ground  state wave function. By making use of the time correlator function, we observe that the localized state exhibits the periodic or quasiperiodic oscillations with time, the delocalized state displays temporal chaos.  We also show that the dynamical feature of the localized and delocalized states remains similar for the condensate trapped in quasiperiodic and random Gaussian disordered potential. One of the main objectives of our work is to obtain a critical value of the non-linearity beyond which the system shows delocalized state. For the quasiperiodic potential, the transition between the localized and the delocalized state can be ascertained on the basis of the localization length, competition between the kinetic and interaction energy etc. However, for random Gaussian potential making a distinction between these states is not so obvious because of the presence of many phases. Therefore, by studying the dynamics of the correlation function of condensate, we provide a robust tool to establish a clear distinction between the localized and the delocalized states.

The structure of our paper is as follows. In Sec.~\ref{sec2}, we provide the governing equations and numerical simulation details. It is followed by a brief description of different quantities, in Sec.~\ref{sec3}, like time correlation function, power spectral density (PSD) and Lyapunov exponent, which we have used to characterize the localization and chaotic dynamics of the delocalized states. In Sec.~\ref{sec4}, we present the results of the numerical simulations on the delocalization in the quasiperiodic optical lattice and random Gaussian disordered potentials. For each type of potential, first, we analyze the effect of the increase in the nonlinear interaction on the ground state of the condensates. Further, we present the dynamics of the condensates using the time correlator function. In Sec.~\ref{sec5}, we conclude our paper.
%The structure of our paper is as follows. In Sec.~\ref{sec2}, we present the governing equation and numerical details of our studies. It is followed by a brief description of different quantities, in Sec.~\ref{sec3}, like time correlation function, power spectral density (PSD) and Lyapunov exponent, which we have used to characterize the localization and chaotic dynamics of the delocalized states. In Sec.~\ref{sec4}, we present the results of the numerical simulations on delocalization first for the quasiperiodic optical lattice followed by the random speckle potential. For each potential, first we analyze the effect of the increase in the nonlinear interaction on the ground state of the condensates. Further, the dynamics of the condensates are presented using the time correlator function. In Sec.~\ref{sec5}, we conclude our paper.

\section{Numerical model}
\label{sec2}

We consider the condensates confined in strong transverse confinement, which can be modeled using the non-dimensional 1D GPEs as~\cite{Adhikari2009}.  
\begin{align}
\mathrm{i} \frac{\partial \psi(x,t)}{\partial t} = \left[ -\frac{1}{2}\frac{\partial^2}{\partial x^2} + V(x) + g \left \lvert \psi(x,t) \right\rvert ^2\right] \psi(x,t),
\label{eqn2}
\end{align}
where,  $V(x)$ is the trapping potential, $g = 2 a_s N/a_\perp^2$ with $a_s$ being the s-wave scattering length is the nonlinearity, $N$ being the total number of atoms in the condensate and $a_{\perp}$ is the length scale corresponding to the transverse harmonic confinement~\cite{Cheng2010}. We have chosen the transverse harmonic length $a_{\perp} = \sqrt{\hbar/(m \omega_{\perp})}$ as the characteristic length scale with $\omega_{\perp}$ being the transverse trap frequency, $\omega^{-1}_{\perp}$ as the characteristic timescale, and $\hbar \omega_{\perp}$ as the characteristic energy scale of the condensate, where $m$ is the mass of an atom, to obtain the non-dimensionalized Eq.~(\ref{eqn2}). The wave function is re-scaled as $\psi(x,t) = a_\perp^{1/2} \tilde{\psi}(x,t)$, where  $\tilde{\psi}(x,t)$ is the nondimensionalized wave function. For brevity, we have omitted the tilde sign over non-dimensionalized wave function. 
%\begin{align}
%\mathrm{i} \hbar \frac{\partial \Psi(x,t)}{\partial t} = %\left[-\frac{\hbar^2}{2m}\frac{\partial^2 }{\partial x^2} + V(x) + %UN\left\lvert\Psi(x,t)\right\rvert^2\right]\Psi(x,t),
%\label{eqn1}
%\end{align}
%where $m$ is the mass of an atom, $N$ the total number of atoms in the condensate, $U = 4\pi \hbar^2a_s/m$ is the nonlinearity with $a_s$ being the s-wave scattering length \tm{and $V(x)$ is the trapping potential}. The normalization condition of the wave function yields $\int d^3r \left|\Psi(x,t)\right|^2 = N$. As we consider transverse harmonic length $a_{\perp} = \sqrt{\hbar/(m \omega_{\perp})}$ as the characteristic length scale, $\omega^{-1}_{\perp}$ as the characteristic time scale, and $\hbar \omega_{\perp}$ as the characteristic energy scale of the condensate, the Eq.~(\ref{eqn1}) can be depicted in the non-dimensional form as follows: 
%Eq.~(\ref{eqn1}) can be written in dimensionless units with the proper choice of $x = \tilde{x}/a_\perp$, $a_\perp = \sqrt{\frac{\hbar}{m\omega}}$, and $\psi(x,;t) = \sqrt{a_\perp^3} \Psi(x;\tau)$ as,

%with normalization $\int_{-\infty}^{\infty}dx|\psi(x,t)|^2 = 1$ and  Here, $\psi(x,t) = a_\perp^{3/2} \Psi(x,t)$. 

In this study, we have separately considered the trapping potential $V(x)$ as quasi-periodic and random Gaussian disordered potential to analyze the characteristics and dynamics of the localization of the condensates.
In experiments, the quasiperiodic potential can be realized as a superposition of two counter-propagating laser beams of slightly  different wavelengths, which takes the form as~\cite{Roati2008}  
%\begin{align}
 %\label{eqn20}
 %V(\hat{z}) = \sum_{i = 1}^2 2s_{i}E_{i}\cos^2(k_i\hat{z})
%%\end{align}
\begin{align}
 \label{eqn3}
 V(x) =  \frac{4\pi^2s_1}{\rm \lambda_1^2}\cos^2\left(\frac{2\pi}{\rm \lambda_1}x\right)+\frac{4\pi^2s_2}{\rm \lambda_2^2}\cos^2\left(\frac{2\pi}{\rm \lambda_2}x\right),
\end{align}
%Here $s_1$, $E_1$ corresponds to primary and $s_2$, $E_2$ corresponds to secondary lattice amplitudes, recoil energies respectively. $E_i = 2\pi^2\hbar^2/(m\hat\rm \lambda_{max}_i^2)$, $k_i = 2\pi/\rm \lambda_{max}_i$ are respective wave numbers. Here $i=1,2$.  
where $s_1$ and $s_2$ denotes the amplitude of the primary and secondary lattice respectively. Following the experimental consideration of the primary and secondary optical lattice wavelengths as $\rm \lambda_1 = 1032 $ nm and $\rm \lambda_2 = 862 $nm~\cite{Roati2008} respectively, we take the ratio of the non-dimensional wave length (in terms of $a_{\perp} \approx 1\mu$m) as $\hat{\rm \lambda}_2/\hat{\rm \lambda}_1\approx 0.86$ for all our simulation runs~\cite{Adhikari2009}.  
%1D bichromatic optical lattice can be  realized as the superposition of two optical lattice potentials generated by two standing-wave of polarized laser beams of slightly different wavelengths and amplitudes~\cite{Roati2008}. 

To understand the resemblance of this localization and associated dynamical behaviour with the ones in the presence of random potential, we have considered the random Gaussian disordered potential consisting of $N_s$ identical spikes randomly distributed along the $x$-axis \cite{Sanchez_Palencia_2008, Sanchez-Palencia2010} with a form
\begin{align}
\label{eqn4}
V(x) = V_0\sum_{j=1}^{N_s} h(x - x_j),
\end{align}
where $V_0$ is the strength of the spike, $h(x - x_j)$ is the potential of the spike at position $x_j$. The spike potential is considered to have the form of Gaussian in space with width $\sigma$ as~\cite{Sanchez_Palencia_2008}
\begin{align}
\label{eqn5}
h(x) = \frac{1}{\sigma \sqrt{\pi}}\exp\left(-\frac{x^2}{\sigma^2}\right).
\end{align}
The auto-correlation of $V(x)$ is denoted as,
\begin{align}
\label{eqn6}
C(d) = \langle V(x)V(x + d) \rangle - \langle V(x) \rangle^2,
\end{align}
where $\left<V(x)\right>$ is the mean of the potential defined as,
\begin{align}
\label{eqn7}
\langle V(x)\rangle\equiv \frac{1}{2L}\int_{-L}^L V(x) dx= \frac{V_0}{D},
\end{align}
with $D$ as the average spacing between the spikes.
Here, the correlation length and correlation energy can be estimated by giving a fit to $C(d)$ (Eq.~\ref{eqn6}) with Gaussian function of the form 
\begin{align}
\label{eqn8}
C(d)\approx V_R^2\exp(-d^2/\sigma_R^2),
\end{align}
where amplitude $V_R$ represents correlation energy and $\sigma_R$ represents the correlation length. To generate the potential we select $N_s = 300$, $L = 30$, and width $\sigma = 0.1$. In this case, the value of correlation energy is $V_R = 4.3974$ and $\sigma_R (\approx \sqrt{2}\sigma) = 0.1340$.

%%%%%%%%%%%%%%%%%%%%%%%%%%%%%%%%%%%%%%%%%%%%%%%%%%%%%%%

\section{Approach to characterize the dynamics of the localized state}
\label{sec3}
In this section, we provide the details of the theoretical approach that we have used to characterize the dynamics of the localized state.

\subsection{Time correlation function and Power spectral density}
For our analysis, we use the time correlation function to characterize the dynamics of the different states. Once we obtain the ground state, we investigate the dynamics of the condensate by a sudden quenching of the nonlinearity of the condensate.
The time correlator function is defined in terms of the absolute of the overlap function as 
\begin{align}
c(t) = \left\lvert\bra{\psi(x,0)} \ket{\psi(x,t)}\right\rvert, 
\label{eqn9}
\end{align}
where $\psi(x,0)$ represents the ground state obtained using imaginary time propagation with respect to repulsive interaction parameters. For convenience we treat this wave function at reference time $t = 0$. After obtaining the ground state $\psi (x,0)$ we quench the nonlinearity to zero in the next time step ($dt$) and evolve the state using the GPE (Eq.~\ref{eqn2}) and obtain the evolved wave function at time $t$ as $\bra{\psi(x,t)}$ with zero nonlinearlity. Here $<.>$ denotes the average over the entire space. The $c(t)$ can be expressed in the more explicit manner as
\begin{align}
c(t)= \left\lvert \int \lvert \psi(x,0)\rvert^2\exp(-i\phi(x, t)) dx \right\rvert
\label{eq:ct}
\end{align}
where, $\phi(x,t)$ is the phase picked up by the ground state wave function $\psi(x,0)$ upon time evolution. The time correlator function can be viewed as the integration of the phase acquired by the quenched state over the whole spatial points at time $t$. There are several laboratory experiments suggest a direct measurement of the phase acquired by the condensate using the atom interferometry~\cite{Shin2004, Saba2005, Shin2005}. Once the phase difference $\phi$ acquired by the quenched condensate is determined the $c(t)$ can be calculated using Eq.~\ref{eq:ct}.

In general, the value of transverse frequency is $\omega_{\perp}=2\pi \times 5$ Hz in a typical experiment ~\cite{Roati2008}. Using this frequency strength, if we convert the real time corresponding to one time step of our simulation which is $dt=5\times 10^{-4}$, we find that it comes out to be $dt\sim 16 \mu s $.  Therefore, the instantaneous quenching time scale in our simulation is of the order of $16 \mu s $.  However, in the laboratory experiments, the quench in the nonlinear interaction is achievable through the Feshbach resonance within a typical time of about $100\, \mu$s~\cite{Chin2010, Ott2003}. We have verified our results by increasing the quenching timescale to $100\, \mu$s considered in the experiment, however, we could not find any significant change in the results presented in this paper.

To get a deeper understanding of the dynamics and, in particular, the chaotic dynamics of the delocalized state, we analyze the power spectral density (PSD) of the $c(t)$, which is given by~\cite{Rao2018,dalui2020},
\begin{align}
\label{eqn13}
\text{PSD} = \dfrac{1}{2\pi \mathcal{N}}\lvert \hat{c}(\omega)\rvert^2, 
\end{align}
where $\hat{c}(\omega)$ is the discrete Fourier transform of the time correlator, $c(t)$ evaluated at $t = m\, dt$ ($m=0,1, \ldots, \mathcal{N}$, and $\mathcal{N}$ is the length of the discrete time series).

\subsection{Lyapunov exponent}
In our studies, we aim to establish a possible connection between delocalization and chaos, which we execute by computing the maximal Lyapunov exponent.
%The quantitative study can give us broad overview of possible connection with chaos and delocalization in the system. localization of states can be verified with the calculation of localization length by exponential fit of the density profile in semilog scale. The increment of localization length with the increasing nonlinearity represents delocalization of states in disorder system. 
In dynamical systems, Lyapunov exponents are defined as the mean rate of divergence of two nearby trajectories with time. In phase space, the rate of divergence of separation between two trajectories, with an infinitesimal initial separation $\delta \bm{X}_0$, can be computed as~\cite{Brezinov2011}, 
%Estimation of Lyapunov exponents for a system governed by dynamical equation solely depends on the initial condition i.e. subtle difference in initial conditions, leads the trajectories to exponentially diverge in nature. Mathematically the definition of the Lyapunov exponent is,} 
\begin{align}
\lvert\delta \bm{X}(t)\rvert \approx \mathrm{e}^{\rm \lambda t}\lvert \delta \bm{X}_0 \rvert
\label{eqn13a}
\end{align}
\begin{align}
\label{eqn14}
\rm \lambda_{max} = \lim_{t\to \infty} \frac{1}{t} \ln \frac{\lvert \delta \bm{X}(t)\rvert}{\lvert \delta \bm{X}_0 \rvert} 
\end{align}
If the exponent is positive ($\rm \lambda_{max} > 0$), neighbouring trajectories diverges exponentially, which is a signature of the chaotic behaviour in the system. 

In general, for the known sets of dynamical equations the exponents can be easily computed by using the phase space trajectory of the variables. However, computation of the exponents directly from the time series is not so straightforward. The reconstruction of phase space from the time series data is mainly executed using the time delay $\tau$~\cite{Vlachos2009} and the embedding dimension $D$~\cite{Kennel1992}. We use Average mutual information (AMI) method to estimate the $\tau$ from the timeseries data, while False nearest neighbour (FNN) method has been used to compute the $D$~\cite{Wallot2018}.
%These parameters are computed with the aid of the Average Mutual Information (AMI) function and the False Nearest Neighbour (FNN)~\cite{Wallot2018}.  
For our analysis, the usual time series of the correlator function $c(t)$ is represented in the phase space using $\tau$ and $D$. Further, we select a reference point on the phase space trajectory, which is used to measure its separation from the nearest neighbour point on the phase space. We store the information of initial separation in $\rm L_0$. As time progresses, the separation between the trajectory is computed until its value exceeds a threshold $\epsilon$. Typically, we consider the $\epsilon\sim \mathcal{O}(10^{-2})$. 
%In this process we store the separation $L_0'$ as  as the threshold value of $\epsilon$ exceeded and continue the process by finding a new nearest neighbour.  
Thereafter, Wolf algorithm~\cite{wolf1985} is used to evaluate the Lyapunov exponents which have three control parameters, namely, $D$, $\tau$, and the threshold value of length $\epsilon$. The Lyapunov exponent is computed using,
\begin{align}
\label{eqn12a}
\rm \lambda_{max} = \frac{1}{N \Delta t} \sum_{i = 1}^{M-1} \log_2 \frac{L^{'}_i}{L_i} 
\end{align}
where, $N$ is the total number of reference points usually depends upon the number of data points present in the timeseries, $\Delta t$ is the time step associated to the time correlator, $\rm L_i$ is the initial separation and $\rm L^{'}_i$ corresponds to the final separation measured for $\rm i^{th}$ segment of the phase space trajectory. For our analysis we consider the embedded dimension $D=3$, time step $\Delta t$ as $0.0005$ for quasiperiodic potential and $0.001$ for random Gaussian disordered potential.

\section{Results}
\label{sec4}
Here, we provide detailed numerical results of ground states and their associated dynamics. For our present analysis, we consider the quasi-periodic potential (Eq.~\ref{eqn3}) and random Gaussian disordered potential (Eq.~ \ref{eqn4}) as two different cases to investigate the localization and quenched dynamics of the localized state by performing the numerical simulation in imaginary- and real-time split-step Crank-Nicolson integration schemes~\cite{Muruganandam2009, YOUNGS2017503}, respectively. In imaginary time propagation, we have considered $dx = 0.025$ and $dt = 0.0005$ for the simulation of the condensate trapped in quasi-periodic potential and for random Gaussian disordered potential, we choose $dx = 0.04$ and $dt = 0.001$.  For all simulation we have used closed boundary condition with $\psi_{boundary}=0$. Also, we have performed the grid independence test by decreasing to $dx= 0.007$ and all the results presented in the paper remain unchanged.  The ground state for different nonlinearity has been obtained using imaginary time propagation scheme in which the Gaussian wave packet centered at $x=0$ has been chosen as an initial condition. 

%The time step is fixed at $dt = 10^{-5}$ for all the simulation runs. 

%The box size $[-30: 30]$ with spatial resolution as  $dx = 0.025$ is chosen for all the simulation runs, and

In the following, we first present our analysis of the condensate trapped in a quasi-periodic optical lattice, and then we will focus on the case of the random lattice. 

\subsection{Delocalization in presence of quasi-periodic optical lattice}
\label{sec:4}

%%%%%%%%%%%%%%%%%%%%%%%%%%%%%%%%%%%%%%%%%%%%%%%%%%%%%%%%%%%%%%%%%%%%%%%%%%
\begin{figure}[!htb]
\includegraphics[width=1.0\linewidth]{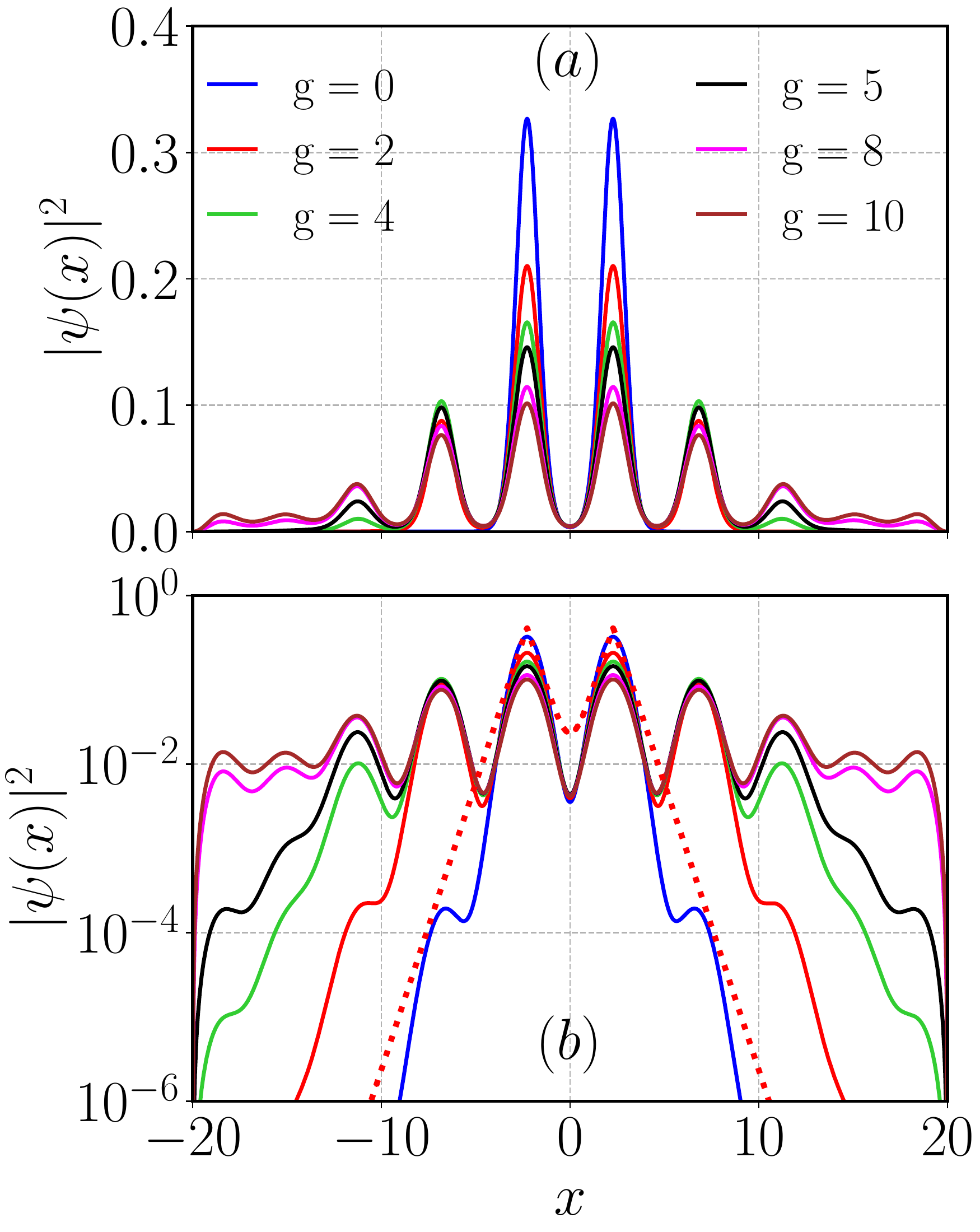}
\caption{Variation of density with different nonlinearity for $\rm \lambda_1 = 10$, $\rm \lambda_2/\rm \lambda_1 = 0.86$, $s_1 = 3$, and $s_2/s_1 = 1.0$ in the (a) linear and  (b) semilog scale. The wave function gets localized near $x=0$ for $g=0$. As we increase the nonlinearity, it results in the spread of the wave function. In the localized states ($g=0, 2, 4$), the condensate density exhibits an exponential tail. For $g \gtrsim 5$, the condensate gets spanned in the whole box, exhibiting delocalized nature. In (b) the dotted red line represents the double exponential fit to the condensate ground state for $g=0$. }
\label{fig1}
\end{figure}%
%%%%%%%%%%%%%%%%%%%%%%%%%%%%%%%%%%%%%%%%%%%%%%%%%%%%%%%%%%%%%%%%%%%%%%%%%%
In this subsection, we first explored the effect of nonlinearity on the localized state of condensates trapped in the bichromatic optical lattice potential. It is followed by an analysis of the dynamical characteristics of the condensates once we switch off the nonlinear interactions after having the ground state of the condensate. As mentioned before, in this case, we vary $\rm \lambda_1$ while keeping the ratio $\rm \lambda_2/\rm \lambda_1 = 0.86$ fixed in Eq.~(\ref{eqn3}). This assumption has been made by following the experimental work of Roati \textit{et al.}~\cite{Roati2008} where the value of transverse harmonic oscillator length is taken as $a_\perp \approx 1\, \mu$m, which yields $ \rm \lambda_1 = 1.032$ and $ \rm \lambda_2 = 0.862$. In our simulations, we consider the space step as $0.025$, while we fix the time step as $0.0005$~\cite{Adhikari2009} and use the Gaussian wave packets centered around zero as an initial condition for all our simulations.

%\sout{
%Following the experimental work of Roati et al.~\cite{Roati2008}, we have considered the parameters for our numerical studies  We consider the value of transverse harmonic oscillator length as $a_\perp \approx 1 \, \mu$m.}\sout{ which yields $\hat \rm \lambda_{max}_1 = 1032$ nm and $\hat \rm \lambda_{max}_2 = 862$ nm. }\sout{In the dimensionless units it becomes $\rm \lambda_{max}_1 \approx 1$ and $\rm \lambda_{max}_2 \approx 0.86$.} 

%%%%%%%%%%%%%%%%%%%%%%%%%%%%%%%%%%%%%%%%%%%%%%%%%%%%%%%%%%%%%%%%%%%%
\begin{figure}
\includegraphics[width=1.0\linewidth]{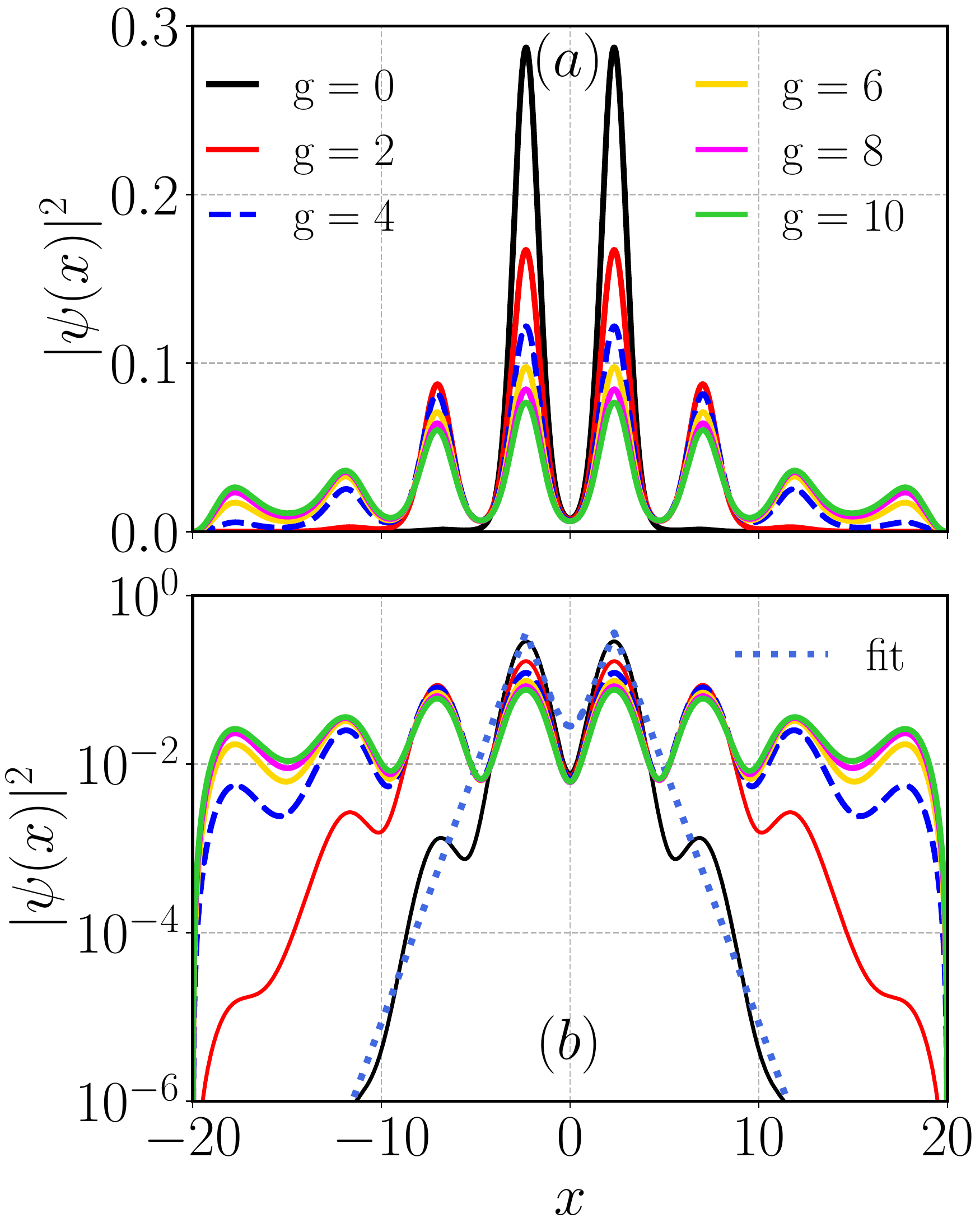}
\caption{(a) Spatial profile of the condensate density for different nonlinearities at  $s_2/s_1 = 0.5$. The other parameters are the same as those in Fig.~\ref{fig1}. The condensate exhibits localized nature for noninteracting cases ($g=0$), which further shows delocalization upon increasing the nonlinearity ($g \gtrsim 4$). (b) Density variation in semilog scale for different nonlinearities. The dotted blue line represents the double exponential fit to the condensate ground state for $g=0$.  The exponential fall of the condensate density around its center characterizes the localization behaviour. The density exhibits delocalized states with increasing nonlinearities. }
\label{fig2}
\end{figure}
%%%%%%%%%%%%%%%%%%%%%%%%%%%%%%%%%%%%%%%%%%%%

%%%%%%%%%%%%%%%%%%%%%%%%%%%%%%%%%%%%%%%%%%%%%%%%%%%%%%%%%%%%%%%%
\begin{figure}
\centering
\includegraphics[width=1.0\linewidth]{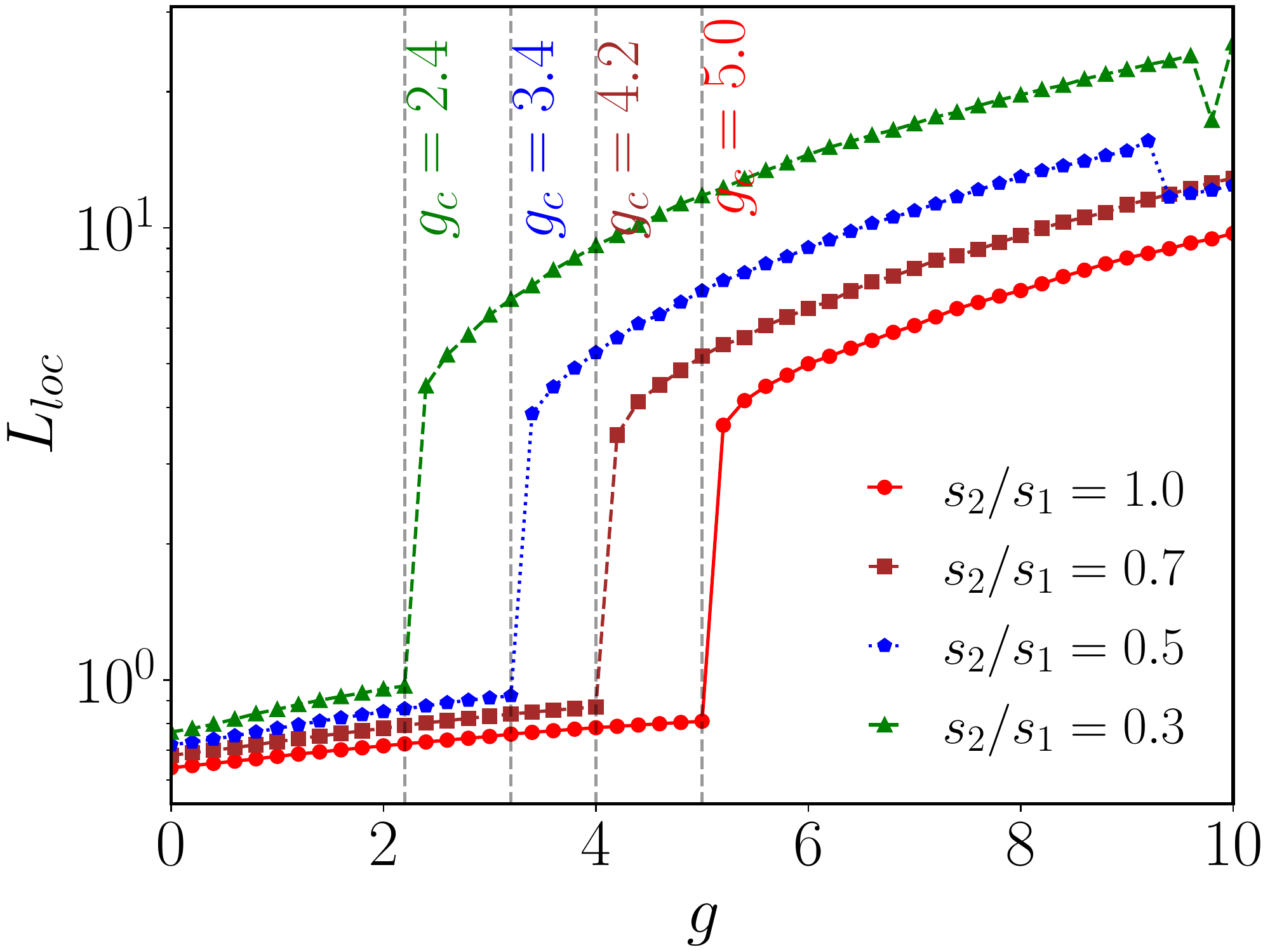}
\caption{Variation of localization length ($L_{loc}$) with the nonlinearity for different $s_2/s_1$. The $L_{loc}$ is calculated using the $1/e$ fall of the fitted curve $a \mathrm{e}^{-\lvert x - x_0\rvert /l }+ a \mathrm{e}^{-\lvert x + x_0 \rvert/l }$ with $x_0$ is point of maximum density in space. Here $x_0 = \pm 2.275$. $L_{loc}$ starts increasing beyond $g_c$. The $g_c$ increases upon increase upon an increase in $s_2/s_1$. The other parameters are the same as those in Fig.~\ref{fig1}.}
\label{fig8}
\end{figure}
%%%%%%%%%%%%%%
\begin{figure*}
\centering
\includegraphics[width=1.0\linewidth]{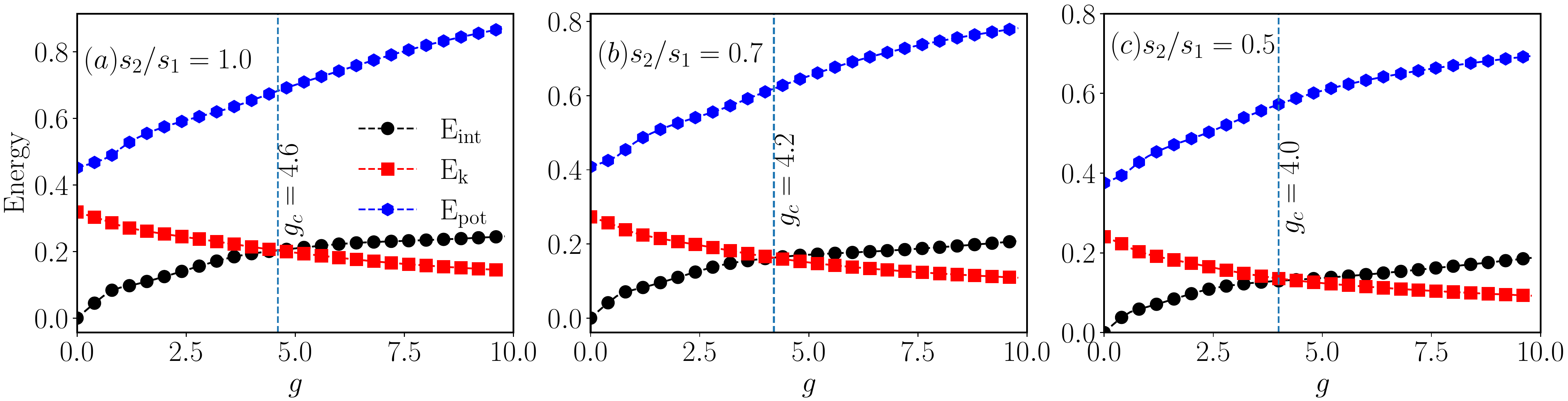}
\caption{Variation of interaction $E_{int}$ ( \protect\markerone ) , kinetic $E_k$ ( \protect\markertwo ) and potential $E_{pot}$ ( \protect\markerthree )  energies with nonlinear interaction for different disorder strength: (a) $s_2/s_1=1.0$, (b) $s_2/s_1=0.7$, and (c)$s_2/s_1=0.5$. All the other parameters are same as Fig.~\ref{fig1} In localized regime $E_k$ dominates over $E_{int}$, however, in the delocalized regime $E_{int}$ dominates over $E_k$. The threshold nonlinear interaction interaction ($g_c$) for different $s_2/s_1$ values are consistent with those obtained from the localization length. The vertical lines are drawn to guide the eyes for $g_c$.}

%\tm{Indicate symbols for different energies. Also increase the symbol sizes a bit.}}
%illustrates the localized and delocalized regions depending upon the crossover of the interaction energy($\rm E_{int}$) over kinetic energy ($E_k$). The threshold value of $g_c$ decreases with the reduction of  disorder amplitude $s_2/s_1$. The potential energy shows the increasing behaviour with $g$. The other parameters are same as Fig.~\ref{fig1}}
\label{fig:en_diff}
\end{figure*}

%%%%%%%%Energy-plot%%%%%%%%%%%
%\begin{figure}
%\centering
%\includegraphics[width=1.0\linewidth]{energy-diff-BOL.pdf}
% \includegraphics[width=1.0\linewidth]{lyp_BOL.eps}
%\caption{Difference between interaction($E_{non}$) and kinetic($E_k$) energy in semilogscale with nonlinearity at different disorder strength $s_2/s_1 = 1.0, 0.7, 0.5$. The threshold value of $g_c$ associated to the minimum value around $E_{int} \approx E_k$ represents the localization to delocalization transition. It is observed that the $g_c$ value decreases with decreasing disorder strength in the system. The other parameters are same as Fig.~\ref{fig1}}
%\label{fig:en_diff}
%\end{figure}

%%%%%%%%%%%%%%%%%%%%%%%%%%%%%%%%%%%%%%%%%%%%%%%%%%%%%%%%%%%
%\subsubsection{Effect of nonlinearity on the localized states}
%\label{sec:4a}
There are two ways through which the delocalization of the condensate trapped in the optical lattice potential can occur. One is to increase the nonlinear repulsive interaction and the other by decreasing the disorder strength upon tuning the ratio of the laser amplitude $s_2/s_1$~\cite{Adhikari2009, Brezinov2011}. We begin by analyzing the effect of nonlinearity on the localized condensates. 
%\tm{The following lines are repeated I think.} \textit{It has been demonstrated that the localization of the weakly interacting condensate exhibits localization, which gets destroyed upon increasing the nonlinearity beyond a certain threshold value~\cite{Adhikari2009}. However, the detailed characterization of the localized and delocalized states is missing. In this paper, we bridge this gap by analyzing the dynamics of these states through an instantaneous quenching of the interaction.} 
%\tm{Unclear statement...Please check!}
The ground states for different localized states for weak nonlinear interaction have already been analyzed by Adhikari and Salasnich~\cite{Adhikari2009}. As we are interested in the analysis of the dynamics of these states, to make the manuscript self-contained, in the following, we briefly present the nature of different ground states for various nonlinearities. 

%\sout{However, in order to make our study self-contained,}\tm{The placement of the following sentence is not appropriate. It should come in the introduction.}\sout{Experimentally, the interaction can be tuned using the Feshbach resonance~\cite{Inouye1998}. }

%%%%%%%%%%%%%%%%%%%%%%%%%%%%%%%%%%%%%
\begin{figure*}[!htp]
\includegraphics[width=0.95\linewidth]{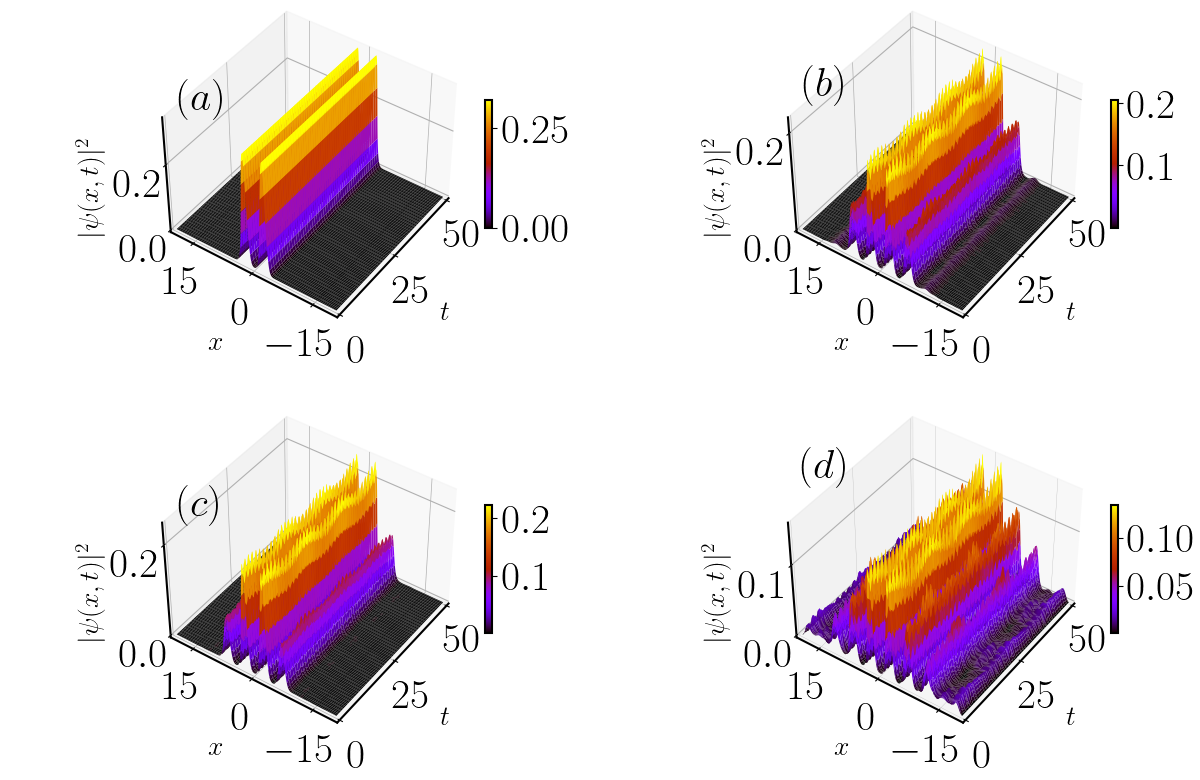}
\caption{Pseudo color representation of the Spatio-temporal evolution of the condensate after quenching the nonlinearity to zero as the ground state was prepared for different $g$: (a) $g=0$, (b) $g=4$, (c) $g=5$, and (d) $g=10$.  Other parameters are same as in Fig.~\ref{fig1}. }
\label{fig3}
\end{figure*}
%%%%%%%%%%%%%%%%%%%%%%%%%%%%%%%%%%%%%%%%%  
  
In Fig.~\ref{fig1}(a), we show the spatial distribution of the ground state density for different nonlinear interactions with $s_1 = 3.0$, $s_2/s_1 = 1.0$, and $\rm \lambda_1 = 10$. It is easy to see that for the non-interacting case ($g = 0$), the condensate gets localized between $-5 \lesssim x \lesssim 5$ with the maximum density around $x \sim \pm 2$. As we increase the nonlinearity to $g=2$, we notice a decrease in the density around the central region ($x\sim0$), resulting in the appearance of peaks at larger values of $x$. However, the condensate appears to be confined between $-10 \lesssim x \lesssim 10$. Further, an increase in the nonlinearity to $g=4$ results in an expansion of the localized condensate in the space. As the nonlinearity exceeds a threshold value ($g \gtrsim 5$), the matter wave localization gets destroyed, which is quite noticeable from the nature of the condensate that appears to get spanned in the whole box as shown for $g=8$ (pink line) and $g=10$ (brown line) in Fig.~\ref{fig1} (a). These features are more noticeable from the behaviour of the tail of the density profile that displays exponential fall in the localized state, a feature which is absent for the delocalized state [cf. Fig.~\ref{fig1}(b)].

For comparison, we also undertake a similar analysis of the localization by lowering the strength of the quasiperiodic optical lattice to $s_2/s_1=0.5$. In Fig.~\ref{fig2}, the spatial profile of the ground state of the matter wave density for different nonlinearities has been illustrated. We find that in this case also, the matter wave remains localized for low nonlinearity as expected. As we increase the nonlinearity, the condensate gets delocalized at a smaller $g$ ($\gtrsim 4$) as compared to those for higher disorder strength ($s_2/s_1 = 1.0$) which happens at $g \gtrsim 5$.   

To quantify this transition, we compute the localization length by fitting the localized states with the function $y = a \mathrm{e}^{-L_{loc}\lvert x - x_0\rvert} + a \mathrm{e}^{-L_{loc} \lvert x + x_0 \rvert}$, where, $L_{loc}$ is the localization length, $a$ is the parameter and $x_0$ is the localization center. The variation of $L_{loc}$ with respect to $g$ is plotted in Fig.~\ref{fig8} for different values of $s_2/s_1$. We find that when the condensate is in the localized state, the $L_{loc}$ varies linearly with $g$. However, a discontinuous jump occurs in $L_{loc}$ for certain value of $g$, beyond which the delocalization happens in the density profile. The threshold value of the nonlinearity at which the delocalization occurs decreases when we decrease the disorder strength $s_2/s_1$. 

%\textcolor{blue}{Further, in order to characterize the phase transition from localization to delocalization we use different energies such as kinetic energy $E_k = \int dx |\frac{\partial \psi}{\partial x}|^2 $, potential energy $E_{pot} = \int dx \hspace{1mm} \psi^{\dag} V(x) \psi $, and the interaction energy $E_{int} = \int dx \hspace{1mm} g \psi^{\dag}  \psi $, where $\psi$ is the ground state obtained using imaginary time propagation of Eq.~\ref{eqn2} with finite $g$. On the other hand, we calculate chemical potential of the stationary state to characterize different phases of the condensate trapped in random Gaussian disordered potential.}

%%% Modified by PM on 11.04.2023 @6:30PM%%
To throw light on the transition of the condensate from the localized to the delocalized state, it is pertinent to include the relevant physical cause. To achieve this, we computed the energies of different components, such as the kinetic energy $E_k = \frac{1}{2}\int dx\lvert\nabla \psi\rvert^2$, potential energy $E_{pot} = \int V(x)\lvert\psi\rvert^2 dx$, and interaction energy $E_{int} = \frac{1}{2}\int g\lvert \psi\rvert^4 dx$, where $\psi$ is the ground state wave function obtained by using imaginary time propagation of Eq.~\ref{eqn2} with finite $g$. In Fig.~\ref{fig:en_diff}, we present the variation of different energies, $E_k$ (\protect\markertwo), $E_{int}$ (\protect\markerone), and $E_{pot}$ (\protect\markerthree), for different $s_2/s_1$. We observe that while $E_{pot}$ and $E_{int}$ increase, $E_k$ decreases with an increase in $g$. Interestingly, we find that below the threshold value of the critical nonlinear strength, ($g_c$), $E_k$ dominates over $E_{int}$, while we observe the opposite trend above the critical $g_c$. This trend holds for all values of $s_2/s_1$ considered.
%It is pertinent to add the relevant Physical cause for the transition of the condensate from the localized to delocalized state. To throw light on this we compute the different component energies, such as, the kinetic energy $E_k=\frac{1}{2}\int dx |\nabla \psi|^2 $, the potential energy $E_{pot} = \int V(x) |\psi|^2 dx$, and the interaction energy $E_{int} = \frac{1}{2}\int g|\psi|^4 dx$, where $\psi$ is the wave function of the ground state obtained using imaginary time propagation of Eq.~\ref{eqn2} with finite $g$.  In Fig.~\ref{fig:en_diff} we plot the variation of different energies, $E_k$ ( \protect\markertwo ), $E_{int}$ ( \protect\markerone ) and $E_{pot}$ ( \protect\markerthree ) for different $s_2/s_1$. We find that while $E_{pot}$ and $E_{int}$ increases, $E_k$ shows the decreasing trend upon increase of $g$. Interestingly, we find that below the threshold value of the critical nonlinear strength ($g_c$), $E_k$ dominates over the $E_{int}$, while the opposite trend is observed above the critical $g_c$. This behaviour is moreover quite true for all values of $s_2/s_1$ considered.

In the following, we will analyze the characteristics of these states using their dynamical evolution.  

\subsubsection{Quench dynamics of the localized and delocalized states}
\label{sec:4b}

To study the detailed dynamics of the localized and delocalized condensate, we consider the ground states obtained for different nonlinearity and perform the time evolution by applying an instantaneous quench of the nonlinear interaction to zero. This protocol introduces the dynamics in the condensates, which have been captured by evolving the GPEs using the real-time scheme~\cite{Muruganandam2009}. 

To probe the spatio-temporal evolution after quenching of the condensate prepared at different values of $g$, in Fig.~\ref{fig3}, we plot the spatio-temporal evolution of the density of the condensates after sudden cessation of $g$. For $g=0$, the localized condensate propagates with time without any distortion as can be seen from Fig.~\ref{fig3}(a). However, the localized condensate at $g=4$ develops oscillatory behaviour, as depicted in Fig.~\ref{fig3}(b). The oscillatory behaviour becomes more and more irregular for higher values of nonlinearity [cf. Fig.~\ref{fig3}(c-d)]. Interestingly, we find that the condensate, which was in the delocalized state exhibits chaotic oscillation with time, as depicted in Fig.~\ref{fig3}(d) (for $g=10$). \%% By PM at 7:35PM
Note that the absence of any spatial change in the condensate after the quench may be attributed to the fact the condensate is trapped tightly near the minima of the external potential, and it remains so with zero nonlinear interaction as time progresses. Also, the dispersion due to the kinetic part is quite low because the trapped energy dominates over other energies for all the values of $s_2/s_1$ (see Fig.~\ref{fig:en_diff}). The role of the random potential here is to contribute to the spatial phase with a random value in the condensate with the time, which also reflects in the temporal behaviour of the time correlator function, which we shall discuss later.
\begin{figure}[!htp]
\centering%
\includegraphics[height=1.0\linewidth]{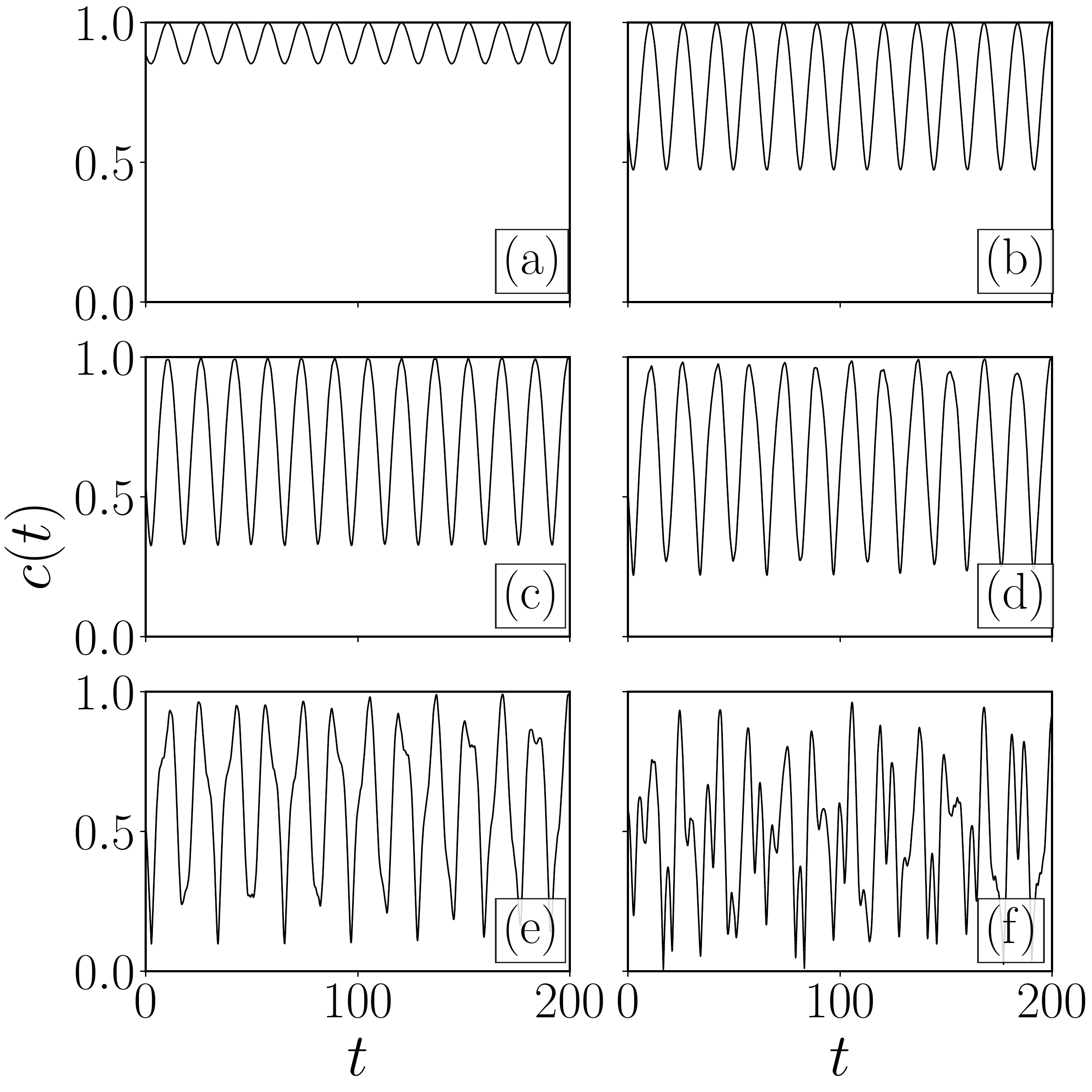}
\caption{Temporal variation of the time correlation function, $c(t)$ at $s_2/s_1 = 1.0$, $s_1 = 3$ and $\rm \lambda_1 = 10.0$ for different nonlinearities (a) $g = 1$, (b) $g = 2$,  (c) $g = 3$, (d) $g = 4$, (e) $g = 5$, and (f) $g = 10$. In the localized state ($g\lesssim 5$), $c(t)$ exhibits periodic  or quasiperiodic behaviour, which becomes chaotic in the delocalized state.  The time period corresponding to periodic oscillation is $T \approx 15.426$ (i.e., $\omega \approx 0.407$).}
\label{fig4}
\end{figure}
%%%%%%%%%%%%%%%%%%%%%%%%%%%%%%%%%%%%%%%%%%%%%%%%%%%%%%%%%%%%%%%%%%%%%%%%%%

We investigate the dynamics of localized matter wave density by computing the time correlator $c(t)$ (\ref{eqn9}) and analyzing its temporal evolution. For this, we consider the ground state obtained for a particular $g$ as $\psi(0)$ and condensate wave function at a time $t$ after quenching of the nonlinearity finite to zero as $\psi(t)$. In Fig.~\ref{fig4}, we show the temporal evolution of $c(t)$ for the ground state prepared at different values of $g$. Figures~\ref{fig4}(a)-(f) depict the evolution of $c(t)$ for $g=1$, $2$, $3$, $4$, $5$, and $10$, respectively. For the localized state prepared at $g = 1$, $c(t)$ exhibits periodic oscillations of amplitude very close to unity. The localized state with $g=2$ shows similar oscillatory behaviour as shown in Fig.~\ref{fig4}(b). For $g = 3$, the $c(t)$ displays some modulated oscillation indicating the presence of more than one frequency. However, for $g=5$ and $g=10$, which correspond to delocalized condensate, the $c(t)$ exhibits aperiodic or chaotic temporal features. This indicates that quenching of nonlinearity generates periodic, quasiperiodic, and chaotic type dynamics depending upon whether the corresponding ground state is localized or delocalized.
%We investigate the dynamics of localized matter wave density by computing the time correlator and analyzing its temporal evolution. The initial state $\psi(0)$ considered in the time correlator function is the ground state function for the given nonlinearity and the $\psi(t)$ is wave function after quenching the nonlinearity of the localized state to be zero. In Fig.~\ref{fig4} we show the temporal evolution of the time $c(t)$ for the localized state prepared at different nonlinearities. Figures~\ref{fig4}(a)-(f) correspond to the $c(t)$ for $g=1,2,3,4,6$ and, $10$ respectively. For the localized state prepared at $g = 1$ we find that $c(t)$ exhibits the periodic oscillation, however the amplitude of the oscillation is very close to one. The localized state with $g=2$ appears to display similar oscillatory behaviour upon quenching the interaction to zero (see Fig.~\ref{fig4}(b)).   However, for $g = 3$ the time correlator displays quasi periodic behaviour. Further, for $g=6$ and $g=10$ which also have delocalized condensate behaviour the correlator appears to exhibit the chaotic behaviour. Overall we find that the quenching of the nonlinearity appears to generate the periodic, quasiperiodic and subsequently chaotic behaviour in the dynamics as the nonlinearity is increased.

%%%%%%%%%%%%%%%%%%%%%%%%%%%%%%%%%%%%%%%%%%%%%%%%%%%%%%%%%%%%%%%%%%%%%%%%%%
\begin{figure}[!htp]
\centering%
\includegraphics[width =1.0\linewidth]{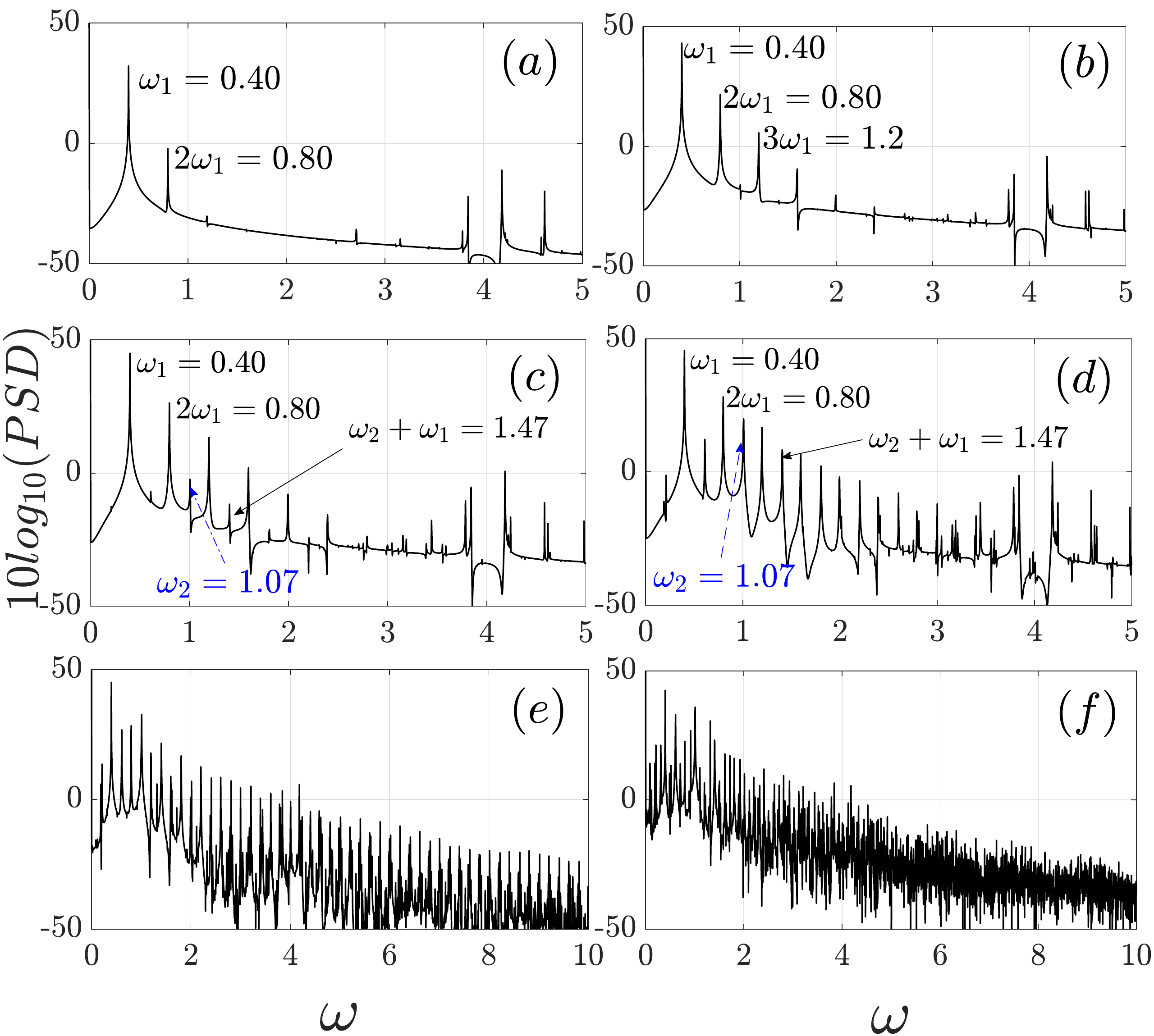}
\caption{PSD of time correlation function (Fig.~\ref{fig2})  for different nonlinearities, (a) $g = 1$ (periodic), (b) $g = 2$ (periodic), (c) $g = 3$ (quasiperiodic), (d) $g = 4$ (quasiperiodic), (e) $g = 5$ (chaotic), and (f) $g = 10$ (chaotic). Other parameters are the same as in Fig.~\ref{fig2}. An increase in the nonlinearity leads to the generation of two incommensurate frequencies $\omega_1=0.40$ and $\omega_2=1.07$ at $g \sim 3$. Finally, a large number of frequencies get generated, a signature of chaotic behaviour at higher nonlinearity ($g\gtrsim 5$).}
\label{fig5}
\end{figure}
%%%%%%%%%%%%%%%%%%%%%%%%%%%%%%%%%%%%%%%%%%%%%%%%%%%%%%%%%%%%%%%%%%%%%%%%%%

To further investigate the nature of the different frequencies present in the dynamics and route to chaotic behaviour of the delocalized state, we compute the PSD of the time correlator, using the formula defined in Eq.~(\ref{eqn13}), corresponding to the periodic, quasi-periodic and chaotic states as discussed above. Figure \ref{fig4} depicts the PSD of $c(t)$  corresponding to the periodic, quasiperiodic and chaotic states. We see that the periodic behaviour of the dynamical state for $g=1$ involves the fundamental frequency $\omega_1=0.40$ together with the presence of its higher harmonics [see Fig.~\ref{fig5}(a)]. We observe similar behaviour in the dynamics of localized state at $g=2$ as depicted in Fig.~\ref{fig5}(b). The PSD of $c(t)$ for $g=3$ shows peaks at the frequencies at $\omega_1=0.40$ and $\omega_2=1.07$, respectively [see Fig.~\ref{fig5}(c)]. The irrational ratio of the two frequencies indicates the quasiperiodic nature of $c(t)$ for $g=3$.  Further, for $g=4$ [Fig.~\ref{fig5}(d)], apart from the frequencies $\omega_1$ and $\omega_2$, other higher frequencies around $\omega_1$ and $\omega_2$ as well as sub-harmonics, like, $\omega_2 + \omega_1$, $\omega_2 + 2\omega_1$, etc. start appearing. We notice more frequencies start getting populated around $\omega_1$ and $\omega_2$ for the dynamics of condensate that show delocalized state for higher nonlinearity (for $g\gtrsim 4$) as depicted in Figs.~\ref{fig5}(e)-(f). The exponential fall behaviour of frequencies in the dynamics of the delocalized state confirms the fully developed temporal chaos~\cite{Sigeti1995, Frisch1981}. The inverse of the rate of decay ($\mu$) of the PSD with the frequency of the chaotic state is of the order of $\sim1.4$ in the high frequency range. In general $\mu$ is related with Lyapunov exponent~\cite{Sigeti1995}. We find that the dynamics of the localized state exhibit periodic oscillations for weak nonlinear interaction, which transforms into quasiperiodic for stronger nonlinear interaction. The value of $g$ at which the condensate exhibits delocalized nature has a chaotic time correlator. With this, we find a systematic generation of the frequencies that finally leads to the chaotic dynamics of the delocalized state which suggests a  quasiperiodic route to chaos. We find the presence of the same quasiperiodic route to chaos for the lower disorder strengths ($s_2/s_1$), which we discuss below. 
\begin{figure}[!htp]
\centering%
\includegraphics[height=1.0\linewidth]{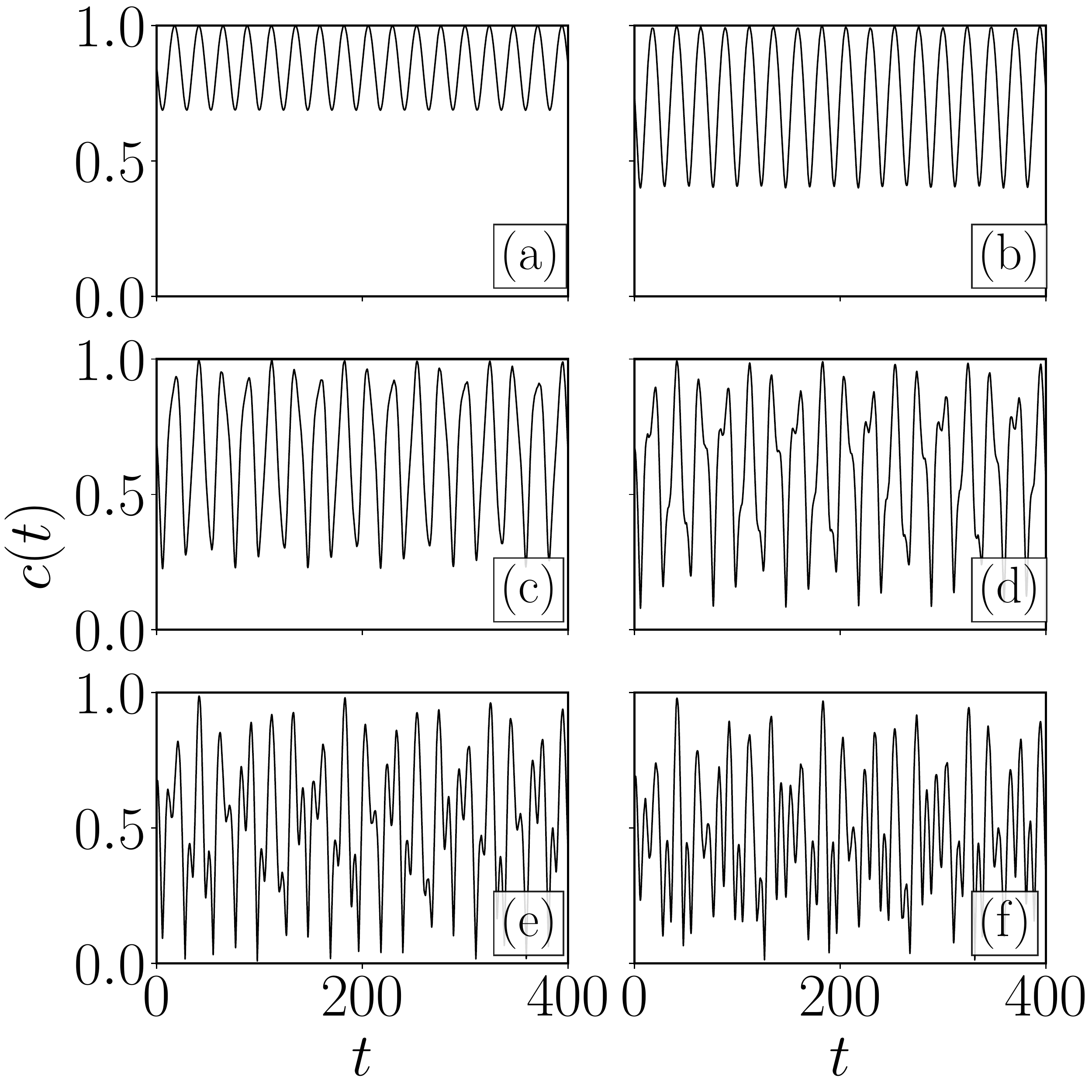}
\caption{Temporal evolution of the time correlator function at $s_2/s_1 = 0.5$, $s_1 = 3$, $\rm \lambda_1 = 10.0$ for different nonlinearities: (a) $g = 1$, (b) $g = 2$,  (c) $g = 3$, (d) $g = 4$, (e) $g = 5$, and (f) $g = 10$. The temporal behaviour of the localization and delocalization behaviour have similar qualitative features as those for ${s_2}/{s_1}=1.0$, and the only difference reflects in the decrease in the nonlinearity at which the correlation shows the chaotic behaviour. The time period for the periodic oscillation is $T \approx 23.691$ ($\omega \approx 0.265$).}
\label{fig6}
\end{figure}
%%%%%%%%%%%%%%%%%%%%%%%%%%%%%%%%%%%%%%%%%%%%%%%%%%%%%%%%%%%%%%%%%%%
  
Next, we move our focus on investigating the nature of the dynamics of the condensate for a lower disorder strength ($s_2/s_1=0.5$). In Fig.~\ref{fig6} we plot $c(t)$ for different values of $g$. Here, the amplitude of $c(t)$ appears to be slightly higher compared to those obtained for $s_2/s_1=1$ (Fig.~\ref{fig4})  after quenching the nonlinearity ($g=1 \rightarrow 0$). Similar to the higher disorder strength, in case of $s_2/s_1 = 0.5$ the $c(t)$ shows periodic (Fig.~\ref{fig6}(a-b)), quasiperiodic (Fig.~\ref{fig6}(c-d)) and chaotic oscillation (Fig.~\ref{fig6}(e-f))  as the nonlinearity is quenched , respectively, from $g=\{1, 2\}$, $g=\{3, 4\}$, and $g=\{5, 10\}$  to zero. One noticeable effect of disorder strength ($s_2/s_1$) is observed in terms of the magnitude of the fundamental ($\omega_1$) and quasiperiodic frequencies ($\omega_1$ and $\omega_2$), which have decreasing trend upon the decrease in $s_2/s_1$.   
%%%%%%%%%%%%%%%%%%%%%%%%%%%%%%%%%%%%%%%%%%%%%%%%%%%%%%%%%%%%%%%%%%%
 \begin{figure}[!htp]
\centering%
\includegraphics[width=1.0\linewidth]{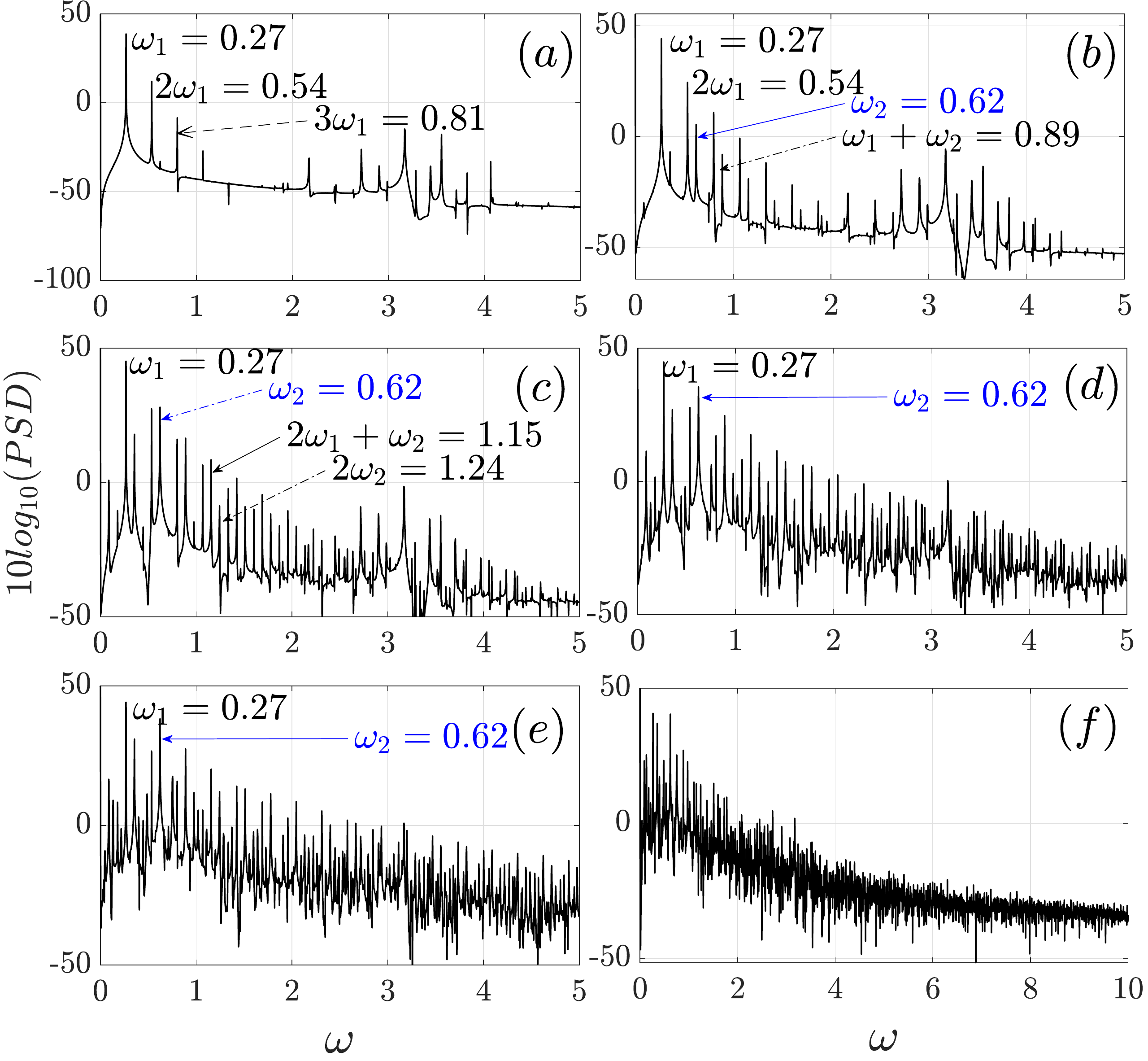}
\caption{PSD of time correlator function shown in Fig.~\ref{fig6} at different nonlinearities: (a) $g = 1$, (b) $g = 2$, (c) $g = 3$, (d) $g = 4$, (e) $g = 5$, and (f) $g = 10$. Other parameters are the same as in  Fig.~\ref{fig5}. The system undergoes a transition from the localized ($g=1, 2, 3$) to the delocalized state ($g=4, 5,10$) with the appearance of frequencies around quasiperiodic frequencies ($\omega_1$ and $\omega_2$) in the PSD of the time correlation function upon increasing the nonlinearity. }
\label{fig7}
\end{figure}
%%%%%%%%%%%%%%%%%%%%%%%%%%%%%%%%%%%%%%%%%%%%%%%%%%%%%%%%%%%%%%%%%%

%%%%%%%%%%%%%%%%%%%%%%%%%%%%%%%%%%%%%%%%%%%%%%%%%%%%%%%%%%%%%%%%%%%%%%%%%%%%%%%%%%
\begin{figure}[!htp]
\centering
\includegraphics[width=1.0\linewidth]{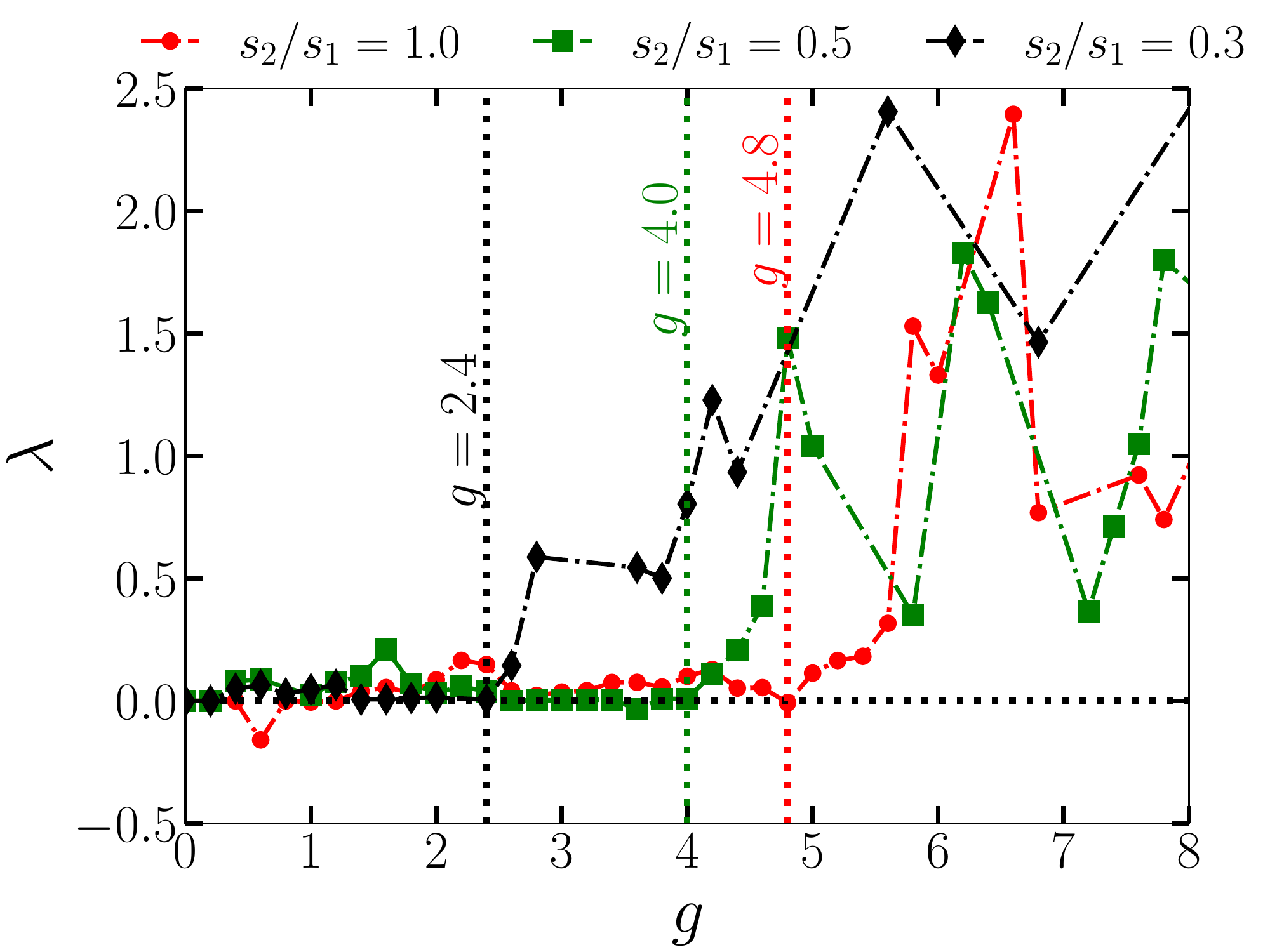}
\caption{Variation of the Lyapunov exponent ($\rm \lambda_{max}$) against nonlinearity for different disordered strengths: $s_2/s_1=1$ (dash-dotted line with solid-red circles), $s_2/s_1=0.5$ (dash-dotted line with solid-green squares) and $s_2/s_1=0.3$ (dash-dotted line with solid black diamonds). For localized state $\rm \lambda_{max} \lesssim 0$, while for delocalized state $\rm \lambda_{max}>0$, indicating the chaotic dynamics. The threshold $g_c$ above which $\rm \lambda_{max}>0$ (a characteristic for the delocalized state) decreases upon the decrease in $s_2/s_1$. Vertical dotted lines are drawn to guide the eyes for different $g_c$. 
}
\label{fig9}
\end{figure}
%%%%%%%%%%%%%%%%%%%%%%%%%%%%%%%%%%%%%%%%%%%%%%%%%%%%%%%%%%%%%%%%%%%%%%%%%%

In Fig.~\ref{fig7}, we illustrate the PSD of the time correlator presented in Fig.~\ref{fig6}. Figure~\ref{fig7}(a) shows the presence of fundamental frequency $\omega_1 = 0.27$ along with its higher harmonics $2 \omega_1 = 0.54$, $3 \omega_1 = 0.81$ in the dynamics of the localized state when the nonlinearity is instantaneously quenched as $g=1 \rightarrow 0$. In the case of quench dynamics of the high nonlinearity state, for example, $g = 2$, we find other eigenfrequencies such as $\omega_2 = 0.62$, apart from the fundamental frequency at $\omega_1 = 0.27$, which indicates the quasi-periodic nature of $c(t)$. At higher nonlinearity $g \gtrsim 4$, more frequencies around $\omega_1$ and $\omega_2$ start getting populated (see Fig.~\ref{fig7}(d)-(f)).  The appearance of other frequencies in the case of the delocalized state ($g=4\rightarrow 0$) exhibits exponential decay behaviour, indicating the presence of chaotic dynamics. We find that for $s_2/s_1 = 0.5$, the chaotic dynamics appear for the state when $g \gtrsim 4$, which is lesser than those for $s_2/s_1 = 1$. However, it is interesting to note that for both the disorder strengths ($s_2/s_1=0.5, 1$), \textit{quasi-periodic route to chaos} have been observed in the dynamics when the condensate makes a transition from the localized to the delocalized state.

%\textcolor{blue}{Attn: Is the ratio of the frequencies $(\omega_1/\omega_2)$ irrational? Please check. Here we claim the nature of the ratio between the frequencies as irrational when its value upto seven of eight decimal places are non-repeating!} 

%%The observation of power spectral density(PSD) confirms the possible chaos phenomenon in energy space with nonlinearity. Next, we confirm the phenomena along with the quantitative view of the chaos. As a consequence, we calculate the lyapunov exponent from time correlation function.
%\begin{figure}
%\centering
%    \includegraphics[width=1.1\linewidth]{fig_7a.eps}
%    \includegraphics[width=1.1\linewidth]{fig_7b.eps}
%    \caption{(a)Variation of density (in semilog scale) at $s_2/s_1 = 1.0$(a), and at $s_2/s_1 = 0.5$ (b) for different nonlinearities. Others parameters corresponding to (a)  and (b) are same as those in Fig.~\ref{fig1} and Fig.~\ref{fig4} respectively. Localization behaviour is characterized by the exponential fall of the condensate density around its center. The density exhibits more spread upon increase in the nonlinearities. }
%    \label{fig8}
%\end{figure}

To further quantify the chaotic nature of the dynamics more systematic manner, we compute the maximal Lyapunov exponent ($\rm \lambda_{max}$) as given in Eq.~(\ref{eqn14}) corresponding to the dynamics of the condensate using the time series of the time correlator function $c(t)$. In Fig.~\ref{fig9}, we show the variation of the $\rm \lambda_{max}$ with the interaction strength for different values of $s_2/s_1$. The increase of $\rm \lambda_{max}$ towards the positive value indicates the chaotic nature of the time correlator function. We find that the Lyapunov exponent fluctuates about zero until $g\sim 4.8$ for $s_2/s_1 = 1$. When $g \gtrsim 4.8$, we witness a systematic increase in the Lyapunov exponent and remains positive ($\rm \lambda_{max}>0$), indicating the chaotic nature of the time correlator functions for those range of $g$. Therefore, the above analysis provides the value of the critical nonlinearity, beyond which the condensate has a delocalized state and thus the corresponding dynamics is chaotic.  Lowering the disorder strength results decrease in the value of the critical nonlinearity ($g_c$) beyond which $\rm \lambda_{max}$ becomes positive. Further, for $s_2/s_1=0.5$, the critical nonlinearity beyond which the condensate has chaotic dynamics is $g_c\sim 4$, while for $s_2/s_1=0.3$ it is $g_c\sim 2.4$. Note that the value of $g_c$ calculated through this analysis provides accurate nonlinearity at which the condensate is delocalized in space and has dynamically chaotic behaviour, which may be important feedback for the experiments. At this juncture, it is worth mentioning that B\v{r}ezinov\'{a} \textit{et al.}~\cite{Brezinov2011} observed a similar kind of chaotic behaviour in the dynamics of the delocalized state when the condensate was subject to weak periodic or aperiodic (quasi-periodic and random Gaussian disordered potential) trap.

 %obtain a systematic dependence of the critical interaction strength on the strength of the optical lattice potential at which chaos occurs.
%We also compute the Lyapunov exponents corresponding to the dynamics of condensate and obtain the systematic dependence of the critical interaction strength on the strength of the optical lattice potential at which chaotic dynamics is observed. In Fig.~\ref{fig9} we plot the variation of the maximum lyapunov with the interaction strength for different $s_1/s_1$. We have used the Wolf's algorithm to compute the Lyapunov exponent from the time series of the time correlator~\cite{wolf1985}. We find that for $g > 4.8$  the value of $\rm \lambda_{max}$ become positive, indicating the chaotic dynamical behaviour of the condensates. Similar type of chaotic nature have been observed in the dynamics of localized condensate by removing the trapping potential after obtaining the ground state for both the quasiperiodic as well the random speckle potential~\cite{Brezinov2011}.

So far, we have analyzed the dynamics of the condensates in the presence of the quasi-periodic potential and found that while the localized state exhibits either periodic or quasiperiodic dynamics, delocalized states show chaotic dynamics upon quenching the nonlinearity to zero. Also, the route to chaos upon increasing the nonlinear interaction appears to be quasi-periodic in nature. Several studies indicate the similarity in the condensate dynamics for the condensate trapped in the quasi-periodic potential and in the random-speckle potential~\cite{Brezinov2011}. To shade more light on this interesting feature, in the following subsection, we present the spatial and temporal behaviour of condensate in the presence of the random Gaussian disordered potential. % Pl. check the red text
%So far we have analyzed the dynamics of the condensates in presence of the quasi-periodic potential and found that the dynamics of the delocalized states are chaotic in nature. Also, the route to chaos upon increasing the nonlinear interaction appears to be quasi-periodic route to chaos in presence of quasi-periodic potential. There are several studies that indicate in the similarity in the behaviour of the condensate in presence of quasi-periodic which is pseudo random and random potential. To test this feature in the next subsection we present the structure and dynamics of condensate in presence of the random speckle potential~\cite{Brezinov2011}.
%%%%%%%%%%%%%%%%%%%%%%%%%%%%%%
%\begin{figure}[!htp]
%\centering%
%\includegraphics[width=\linewidth]{den-sub.pdf}
%\caption{Variation of density for the condensate trapped in the random speckle potential of strength $V_0=1$ for various nonlinearities: (a) $g = 0$, (b) $g = 1$, (c) $g = 3$, (d) $g = 5$, (e) $g = 8$, and (f) $g =  10$. On increasing the nonlinearity, a reduction of density near $x = 0$ accompanied by new density peaks formed between $[-30, 30]$ is observed. }
%\label{fig11}
%\end{figure}
%%%%%%%%%%%%%%%%%%%%%%%%%%%%%%%%%%
\subsection{Delocalization in presence of random Gaussian disordered potential}
\label{subsec:B}
In this subsection, we discuss the effect of the non-linearity on the localized condensate trapped in the random Gaussian disordered potential. The details to generate the random potential are given in Sec.~\ref{sec2}. First, we discuss the ground states of the condensate trapped in the random Gaussian disordered potential at different nonlinearities for single disorder realization.  It is followed by the dynamics of the condensate in similar line of analysis performed for the condensate trapped in quasiperiodic potential, where we have used the quenching of nonlinear interaction from finite to zero to generate the dynamics. Finally, we characterize the dynamics using the PSD and largest Lyapunov exponent analysis of the time correlator function.

%\textcolor{blue}{We checked the stationary states with different disorder realizations and found similar phase transition of the localized condensates, except the center of the localized states can be different}.
%We use the built-in Fortran routine to generate a random-disordered lattice by mapping the random numbers into an interval $[-L, L]$ through linear transformation and denote the positions ($x_j$) as stated in Eq.~(\ref{eqn4}). We choose the parameters $M = 300$, $L = 30$ ~\cite{Cheng2010} and a small width $\sigma = 0.1$.
%To generate random disordered lattice we use the function RAND in FORTRAN and the random numbers mapped into an interval $[-L,L]$ by a linear transformation and denote the positions ($x_j$) as stated in Eq.~(\ref{eqn4}). We choose the parameters $M = 300$, $L = 30$ ~\cite{Cheng2010} and a small width $\sigma = 0.1$.

%%%%%%%%%%%%%%%%%%%%%%%%%%%%%%%%%%%%%%%%%%%%%%%%%%%%%%%%%%%%%%%%%%%%%%%%
\begin{figure}[!htp]
\centering%
\includegraphics[width=\linewidth]{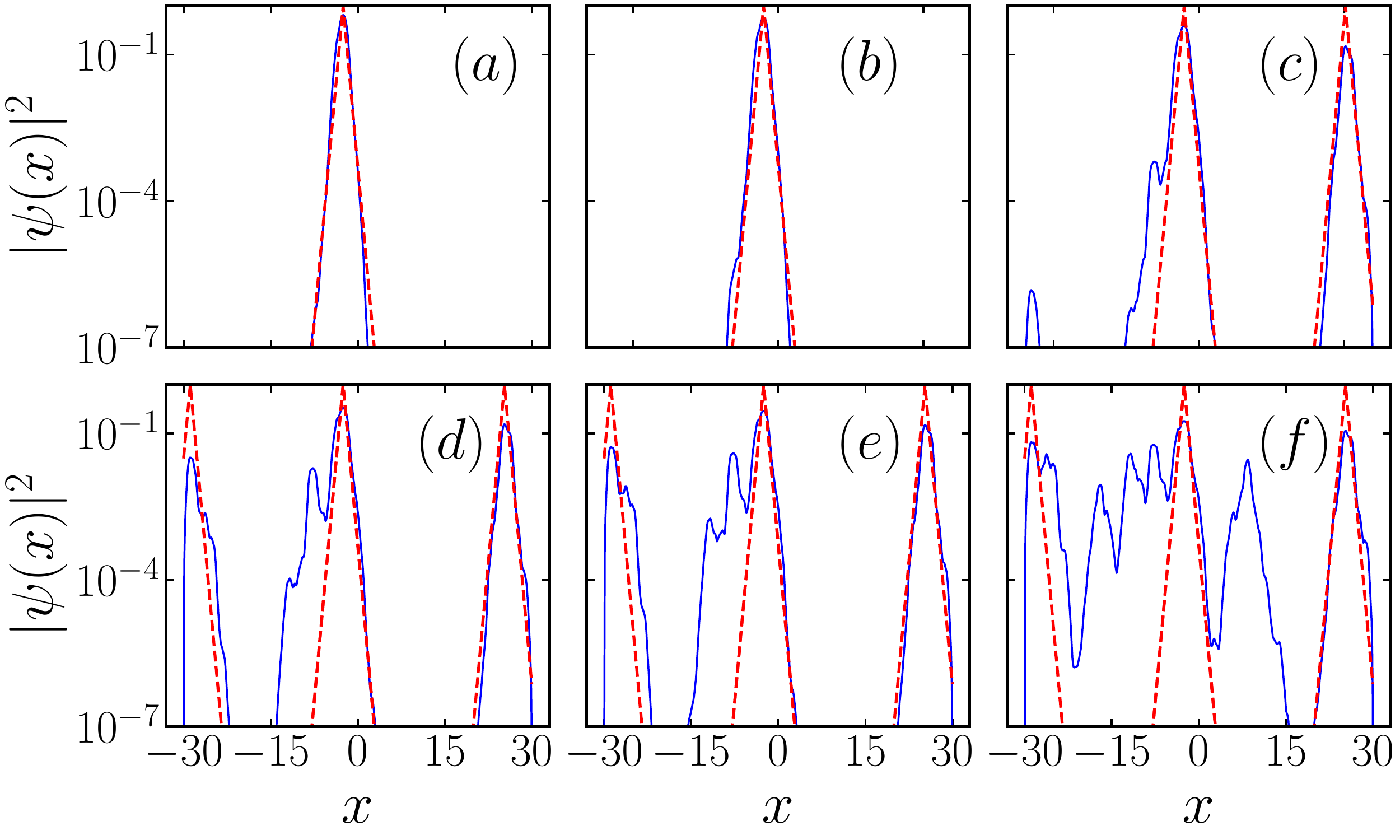}
\caption{Variation of density (in semilog scale) at $V_0 = 1.0$ for different nonlinearities: (a) $g = 0$, (b) $g = 1$, (c) $g = 3$, (d) $g = 4$, (e) $g = 5$, and (f) $g =  10$. The dotted red line is the exponential function drawn near the localized region for the guide of the eyes to show the deviation of the matter wave from the localized nature.}
\label{fig11}
\end{figure}
Unlike the quasi-periodic potential, for this case the condensate exhibits different kinds of phases upon increasing $g$, namely the presence of localized state~\cite{lifshits1988}, Bose glass, and delocalized state of the condensate~\cite{Fisher1989, Scalettar1991, Stellin2022}.  Lugan \textit{et al.} theoretically demonstrated that the condensate undergoes a transition from Lifshitz glass to a delocalized state upon the increase of the nonlinear interaction~\cite{Lugan2007}. In Fig.~\ref{fig11}, we plot the density profile (blue line) in a semi-log scale for different values of $g$. Here, the red dashed lines are the exponential curve near the localized region drawn to display an estimate of the exponential fall of the density in space. For the non-interacting condensate, i.e., $g = 0$, the density profile shows the exponential fall, which is quite evident from the excellent matching of the density profile with the drawn exponential curve [see Fig.~\ref{fig13}(a)] complementing the localized nature of the condensate. The spatial profile of the condensate for $g=1$ also shows exponential fall as depicted in Fig.~\ref{fig13}(b). However, as discussed earlier, due to the random nature of the potential, we witness the presence of another region of the localized condensate, apart from what is present near $x\sim 0$, on increasing $g$ further. For instance, at $g=3$, one part of the condensate gets localized near $x=0$, while another part gets localized near $x \sim 25$. The condensate near both regions appears to fall exponentially, which is quite clear from the fitting of the spatial profile of the condensate with the exponential curve (red dotted line). Note that such kind of bifurcations of the condensate into multiple localized states has been in general termed as fragmented BECs~\cite{Cheng2010, Santos2021} or Bose glass~\cite{Fisher1989, Scalettar1991, Stellin2022}. However, on further increase in $g$ ($ \gtrsim 5$) results in the deviation of the tail of the localized condensate from the exponential nature, as apparent from Fig.~\ref{fig11}(d)-(f) also termed as the delocalized state.

\begin{figure}[!htp]
\includegraphics[width=0.95\linewidth]{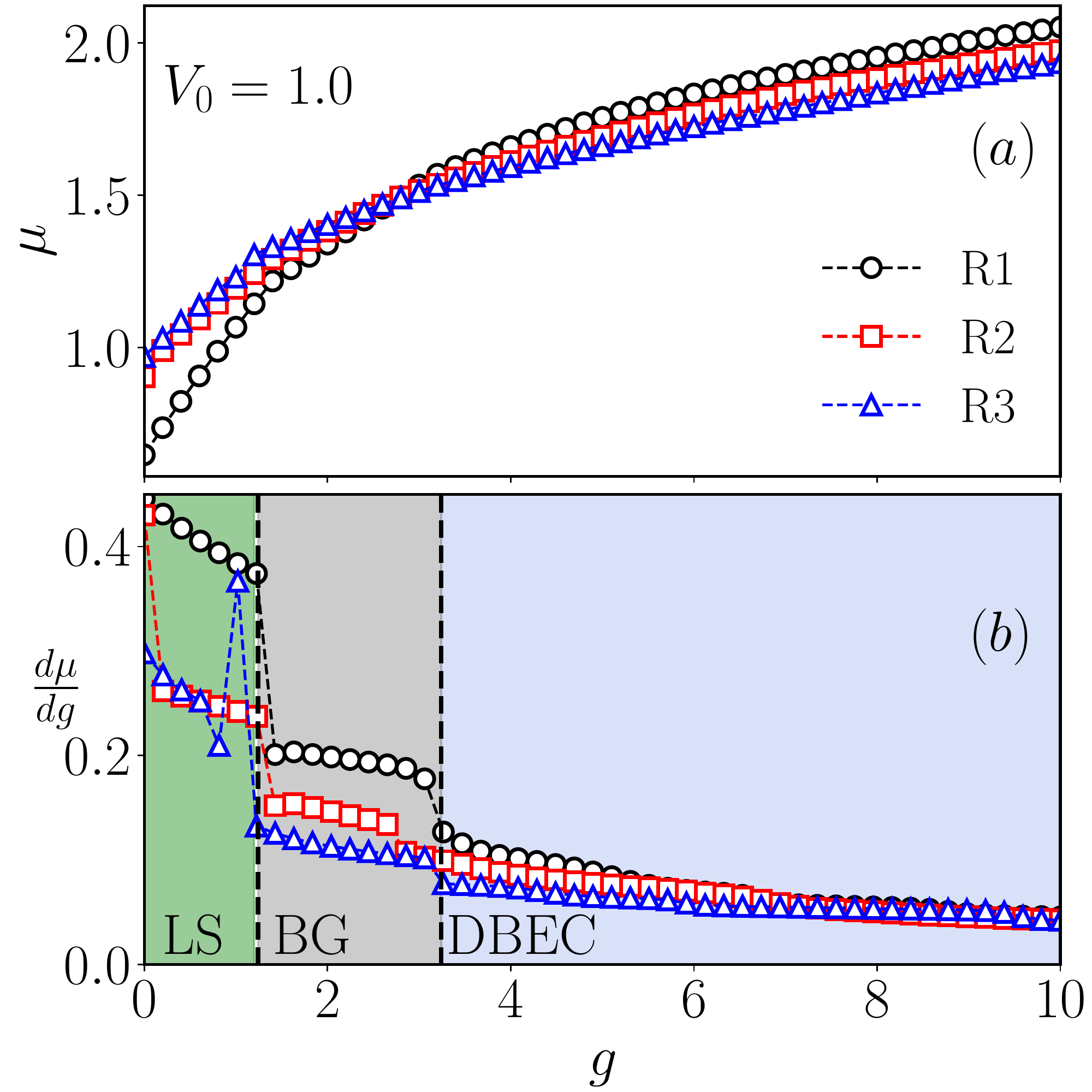}
\caption{Plots showing the different states of the condensate trapped in the random disordered Gaussian potential of strength $V_0=1.0$ based on the nature of $\mu$ in different ranges of $g$. The other parameters are the same as Fig.~\ref{fig11}. (a) Variation of chemical potential $\mu$ with $g$ for three different random realizations: $R_1$ (black circle), $R_2$ (red square), and $R_3$ (blue triangle) (b) Variation of $d\mu/dg$ with $g$ distinguishing different regimes for the condensate: Localized state (LS) for lower $g$ with higher $d\mu/dg\sim 0.2-0.4 $, Bose glass (BG) state at intermediate $g$ with $d\mu/dg\sim 0.12-0.2$, and delocalized BEC (DBEC) state at higher $g$ with $d\mu/dg\sim 0.05 - 0.12$.  The vertical dotted lines are drawn to guide the eyes to distinguish different localized states of the condensate.}

%(a) Variation of chemical potential $\mu$ with nonlinear interaction $g$ for three different random realizations, (b) Differential change of chemical potential with  $g$ which exhibits three different regions, namely Localized state (LS), Bose glass (BG), and delocalized BEC (DBEC) state decided over the values of $d\mu/dg$.}
\label{fig:chem-V1}
\end{figure}

\begin{figure}[!htp]
\includegraphics[width=1.0\linewidth]{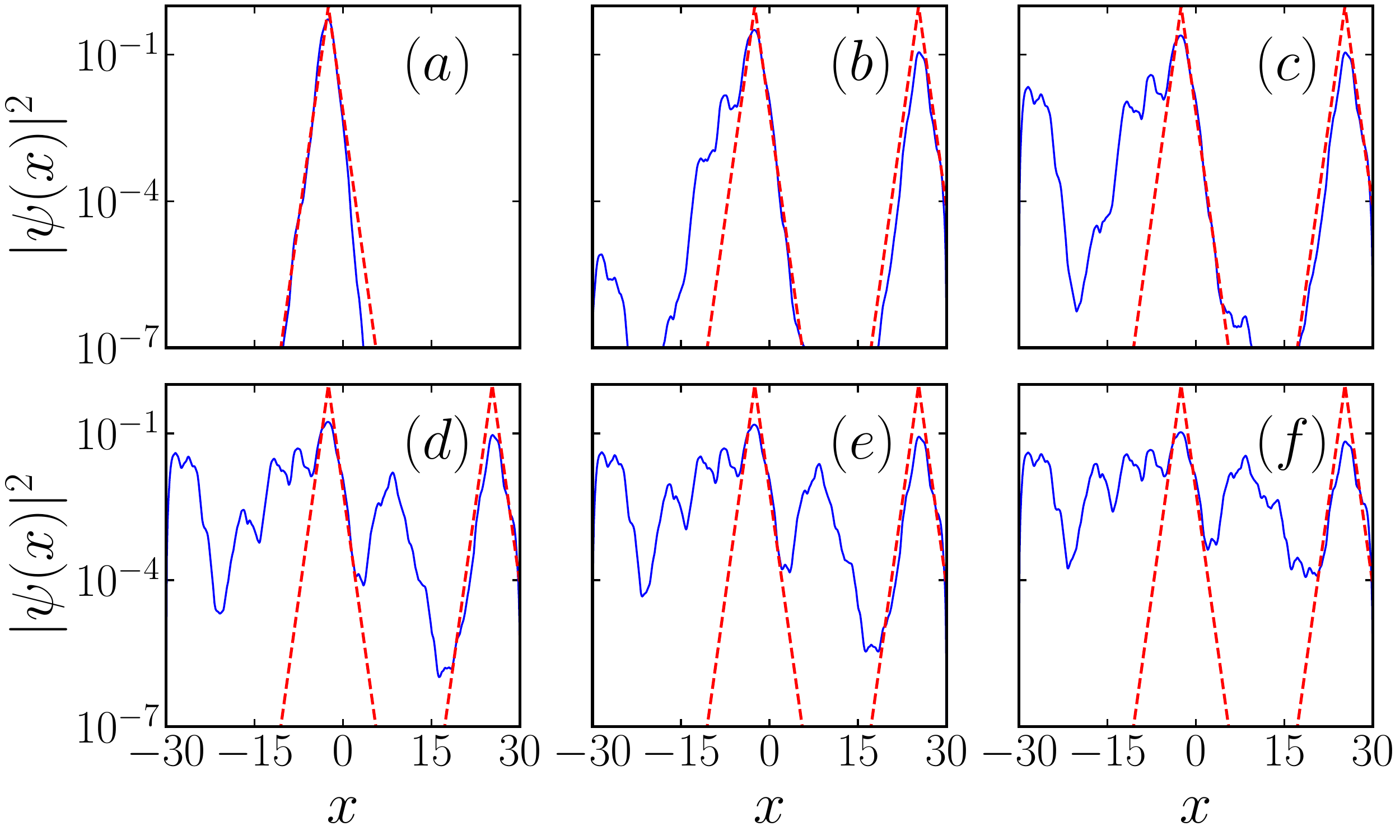}
\caption{Variation of density (in semilog scale) trapped in random potential at $V_0 = 0.5$ with different nonlinearities: (a) $g = 0$, (b) $g = 2$, (c) $g = 3$, (d) $g = 5$, (e) $g = 6$, and (f) $g =  10$. The dotted red line is the exponential function drawn near the localized region for the guide of the eyes to show the deviation of the matter wave from the localized nature.} 
\label{fig13}
\end{figure}

 To make the different regimes of the condensates more explicit, in Fig.~\ref{fig:chem-V1}(a), we plot the $\mu$ as a function of $g$ for three different random realizations ($R_1, R_2$, and $R_3$). We find the presence of three different regimes, which can be identified based upon the values of $d\mu/dg$ for different $g$ as shown in Fig.~\ref{fig:chem-V1} (b). In the localized state (LS), $d\mu/dg$ have the largest value, the intermediate value of $d\mu/dg$ represents the Bose glass (BG), while the lowest $d\mu/dg$  represents the delocalized state (DS). This feature is consistent with the earlier studies~\cite{Fisher1989, Scalettar1991, Stellin2022}.

%%%%%%%%%%%%%%%%%%%%%%%%%%%%%%%%%%%%%%%%%%%%%%%%%%%%%%%%%%%%%%%%%%%%%%%%%%

%%%%%%%%%%%%%%%%%%%
%%%%%%%%%%%%%%%%%%%%%%%%%%%%%%%%%%%%%%%%%%%%%%%%%%%%%%%%%%%%%
%plot to chemical potential
\begin{figure}[!htp]
\includegraphics[width=0.95\linewidth]{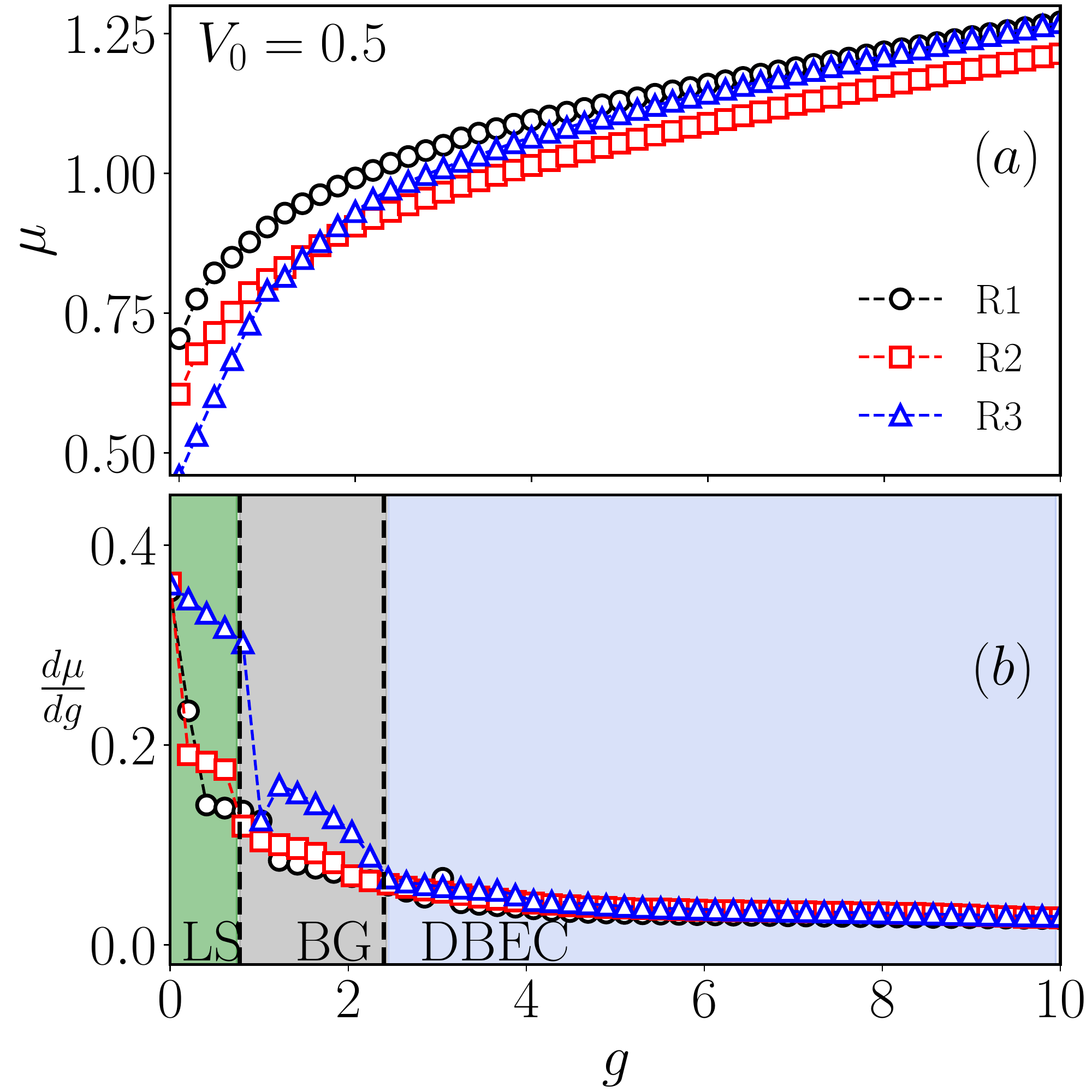}
\caption{Plots showing the different states of the condensate trapped in the random Gaussian potential of strength $V_0=0.5$ based on the nature of $\mu$ in different ranges of $g$. The other parameters are the same as Fig.~\ref{fig13}. (a) Variation of chemical potential $\mu$ with nonlinear interaction $g$ for three different random realizations: $R_1$ (black circle), $R_2$ (red square), and $R_3$ (blue triangle) (b) Variation of $d\mu/dg$ with $g$ distinguishing different regimes for the condensate: Localized state (LS) for lower $g$ with higher $d\mu/dg\sim 0.12-0.36 $, Bose glass (BG) state at intermediate $g$ with $d\mu/dg\sim 0.06-0.12$, and delocalized BEC (DBEC) state at higher $g$ with $d\mu/dg\sim 0.04-0.06$.  The vertical dotted lines are drawn to guide the eyes to distinguish different localized states of the condensate.}
\label{fig:chem-V0p5}
\end{figure}

%%%%%%%%%%%%%%%%%%%%%%%%%%%%%%%%%%%%%%%%%%%%%%%%%%%%%%%%%%%%%%
\begin{figure*}[!htb]
\centering
\includegraphics[width=0.95\linewidth]{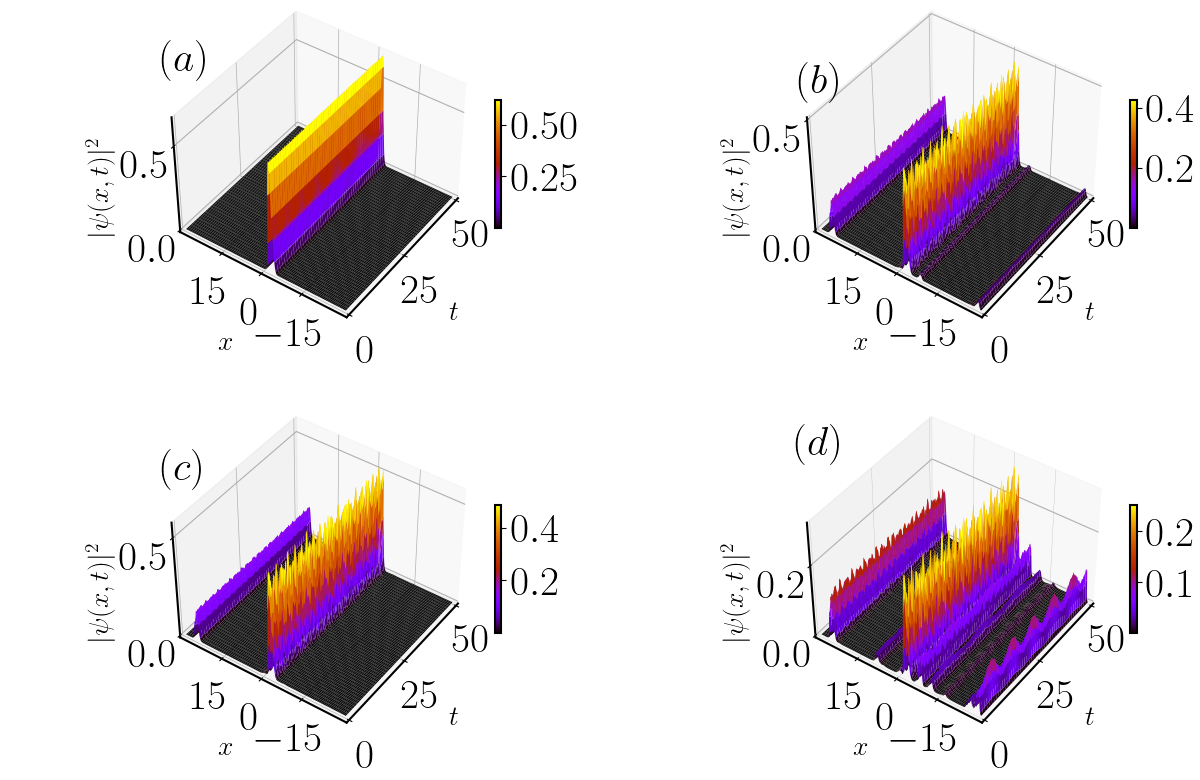}
\caption{Spatiotemporal evolution of the condensate at different nonlinearities (a)$g = 0$, (b) $g = 4$, (c) $g = 5$, and (d) $g = 10$ trapped under the  Gaussian random-disordered potential for $V_0 = 1.0$. The condensate remains localized near $x=-2.51$ for low non-linearity $g=1$. An increase in the nonlinearity results in the delocalized condensate along with more fluctuations in the density, especially near $x=0$.}
\label{fig14}
\end{figure*}

In Fig.~\ref{fig13}, we show the density profile in the semi-log scale for $V_0=0.5$ of the random Gaussian potential. For this parameter, we find that for the lower nonlinearity, ($g \lesssim 3$) the condensate is localized and we observe a delocalized state for higher nonlinearities. However, all the phases, like, LS, BG, and DBEC are of similar nature as those for $V_0=1$ [see Fig.~\ref{fig11}]. It is straightforward to see that decreasing the potential strength decreases the critical nonlinearity above which the condensate gets delocalized, which is around $g_c \sim 3$ for $V_0=0.5$ compared to that for $V_0=1$.
To further identify the different regions of the condensate upon increasing the nonlinearity, In Fig.~\ref{fig:chem-V0p5} we show the variation of $\mu$ as well $d\mu/dg$ with $g$ for $V_0=0.5$ in the panel (a) and (b) respectively. For this case also we also obtain the presence of different localization regimes based on the values of $d\mu/dg$ as those observed for $V_0=1.0$. Note that although the $\mu$ is lower for $V_0=0.5$ than those for $V_0=1$, the value of  $d\mu/dg$ for different regimes of localization appears to be of the same values.

In the following, we present the dynamics of the localized and delocalized state.
%%%%%%%%%%%%%%%%%%%%%%%%%

\subsubsection{Quench dynamics of the condensates trapped in random potential}
\label{sec:B1}
%\textcolor{blue}{In Fig.~\ref{fig14}, we illustrate the temporal evolution of the condensates after quench the nonlinearity to zero from different initial $g$. For $g = 1$ [cf. Fig.~\ref{fig14}(a)], the localized condensate exhibits tiny oscillations with time. Interestingly, for higher non-linearities  [see Fig.~\ref{fig14}(b)-(d]) the condensate becomes fragmented and localized at different lattice sites. However, the temporal oscillation becomes more irregular for higher nonlinearity, and the fragmented condensates tending to expand along $x$ direction. However, the disordered potential obstruct the expansion of the condensates. Subsequently, at high repulsive interaction, beyond which the condensate gets delocalized, the dynamical evolution shows irregular behaviour, and associated dynamics display chaos, a dynamical feature similar to that obtained for the condensate trapped in quasi-periodic potential as studied in Sec.~\ref{sec:4b}. }             

In Fig.~\ref{fig14}, we depict the spatio-temporal evolution of the condensate density after quenching of the nonlinearity to zero from different initial values of $g$. For $g=0$ [cf. Fig.~\ref{fig14}(a)], the localized condensate propagates with time without any distortion. The condensate develops fluctuation with time upon the increase in the value of $g$ [see Fig.~\ref{fig14}(b)-(d]), especially near the region $x=0$. The temporal oscillation becomes more irregular for higher nonlinearity, and the corresponding dynamics display chaos, a dynamical feature similar to that obtained for the condensate trapped in the quasi-periodic potential. Note that as discussed for the situation of no expansion along the x-direction for the condensate trapped in quasi-periodic potential once the dynamics appear due to quenching in the non-linearity, we find a similar scenario for the condensate dynamics trapped with random potential. In this case, we also observed that the potential energy dominates over the kinetic and thus does not allow the condensate to diffuse around the minima of the potential well.
%%%%%%%%%%%%%%%%%%%%%%%%%%%%%%%%%%%%%%%%%%%%%%%%%%%%%%%%%%%%%%%%%%%%%
%\newpage
\begin{figure}[!htb]
\centering%
\includegraphics[height=1.0\linewidth]{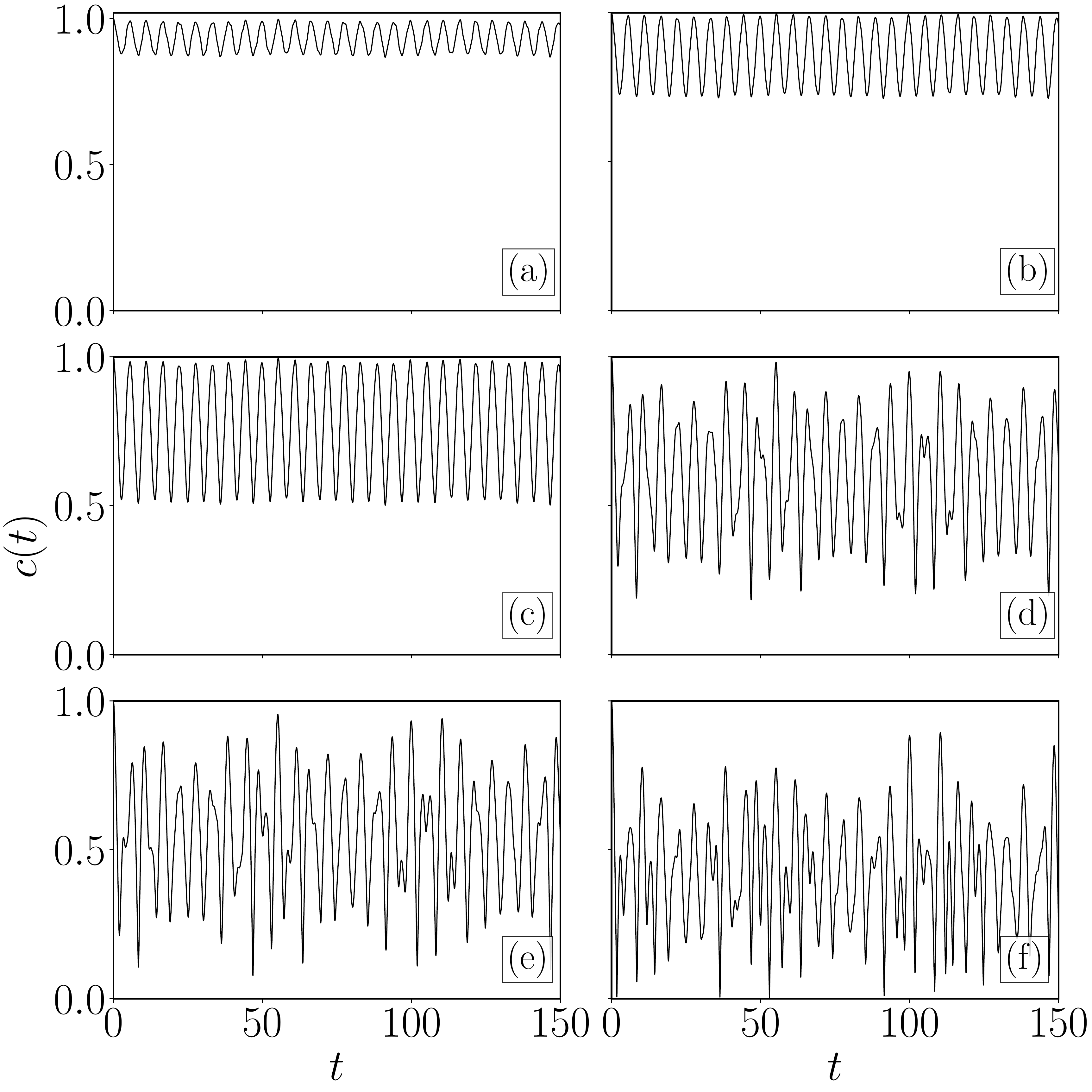}
\caption{Temporal evolution of the time correlation function for quenching from (a) $g = 1.6$, (b) $g = 2$, (c) $g = 3$, (d) $g = 5$, (e) $g = 6$, and (f) $g = 10$ to $g=0$. The other parameters are the same as in Fig.~\ref{fig11}. In the localized state($g \lesssim 5$), the correlation function exhibits a periodic or quasiperiodic, which becomes chaotic at higher $g$.}
\label{fig15}
\end{figure}
%%%%%%%%%%%%%%%%%%%%%%%%%%%%%%%%%%%%%%%%%%%%%%%%%%%%%%%%%%%%%%%%%%%%%
%%%%%%%%%%%%%%%%%%%%%%%%%%%%%%%%%%%%%%%%%%%%%%%%%%%%%%%%%%%%%%%%%%%%%%%%%%
\begin{figure}[!htb]
\centering%
\includegraphics[width=1.0\linewidth]{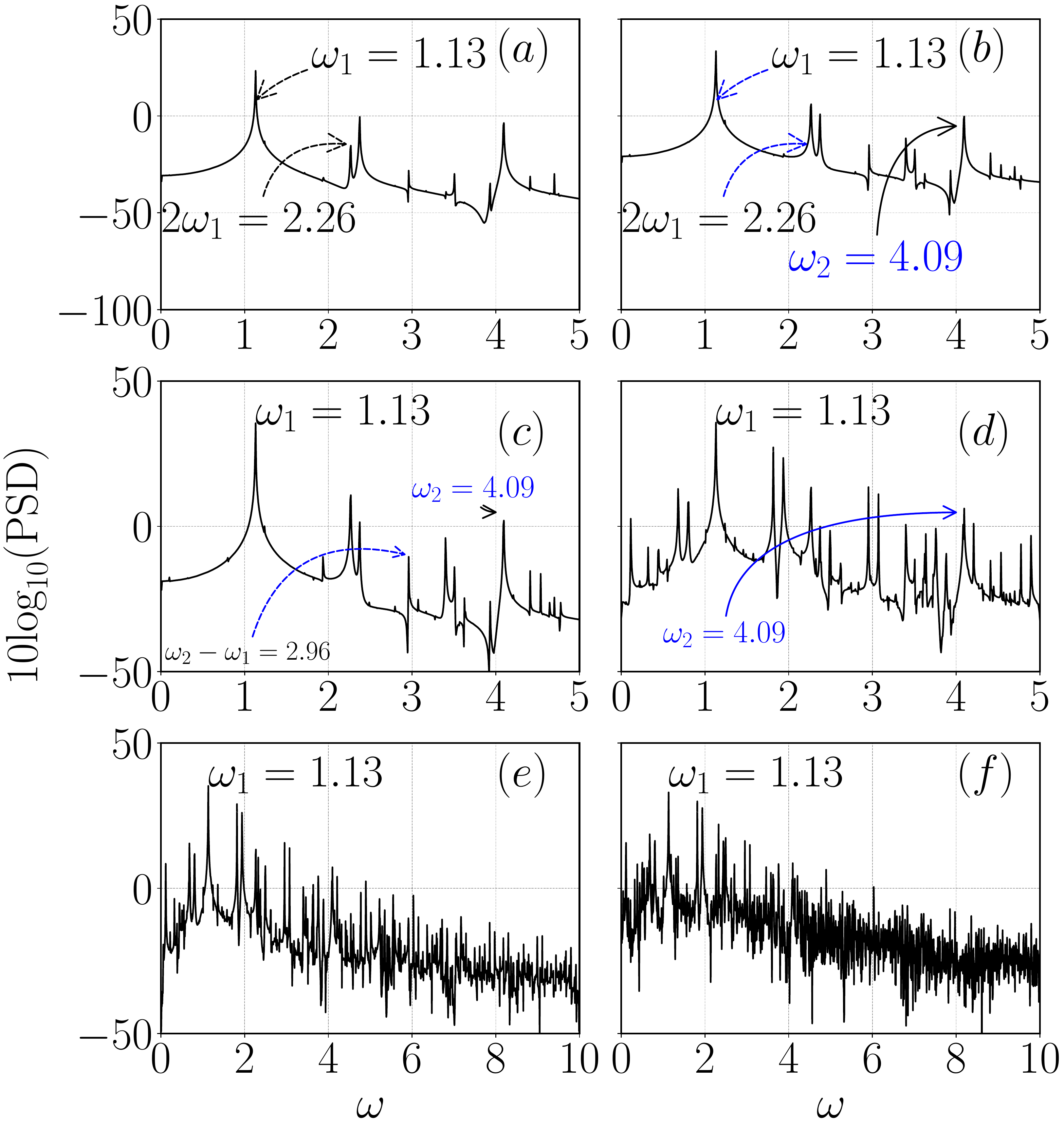}
\caption{PSD of time correlation function (as shown in Fig.~\ref{fig15}) for the situation when the nonlinearity is quenched from (a) $g = 1.6$, (b) $g = 2$, (c) $g = 3$, (d) $g = 5$, (e) $g = 6$, and (f) $g =10$ to $g=0$ for $V_0=1$. Increasing the nonlinearity generates two incommensurate frequencies $\omega_1 = 1.13$, and $\omega_2 = 4.09$ at $g \sim 2$. Finally, the region near these frequencies starts getting populated, leading to the chaotic behaviour at higher nonlinearity ($g\gtrsim5$). }
\label{fig16}
\end{figure}
%%%%%%%%%%%%%%%%%%%%%%%%%%%%%%%%%%%%%%%%%%%%%%%%%%%%%%%%%%%%%%%%%%%%%%%%%%
%\newpage
%%%%%%%%%%%%%%%%%%%%%%%%%%%%%%%%%%%%%%%%%%%%%%%%%%%%%%%
\begin{figure}[!htb]
\includegraphics[width=\linewidth]{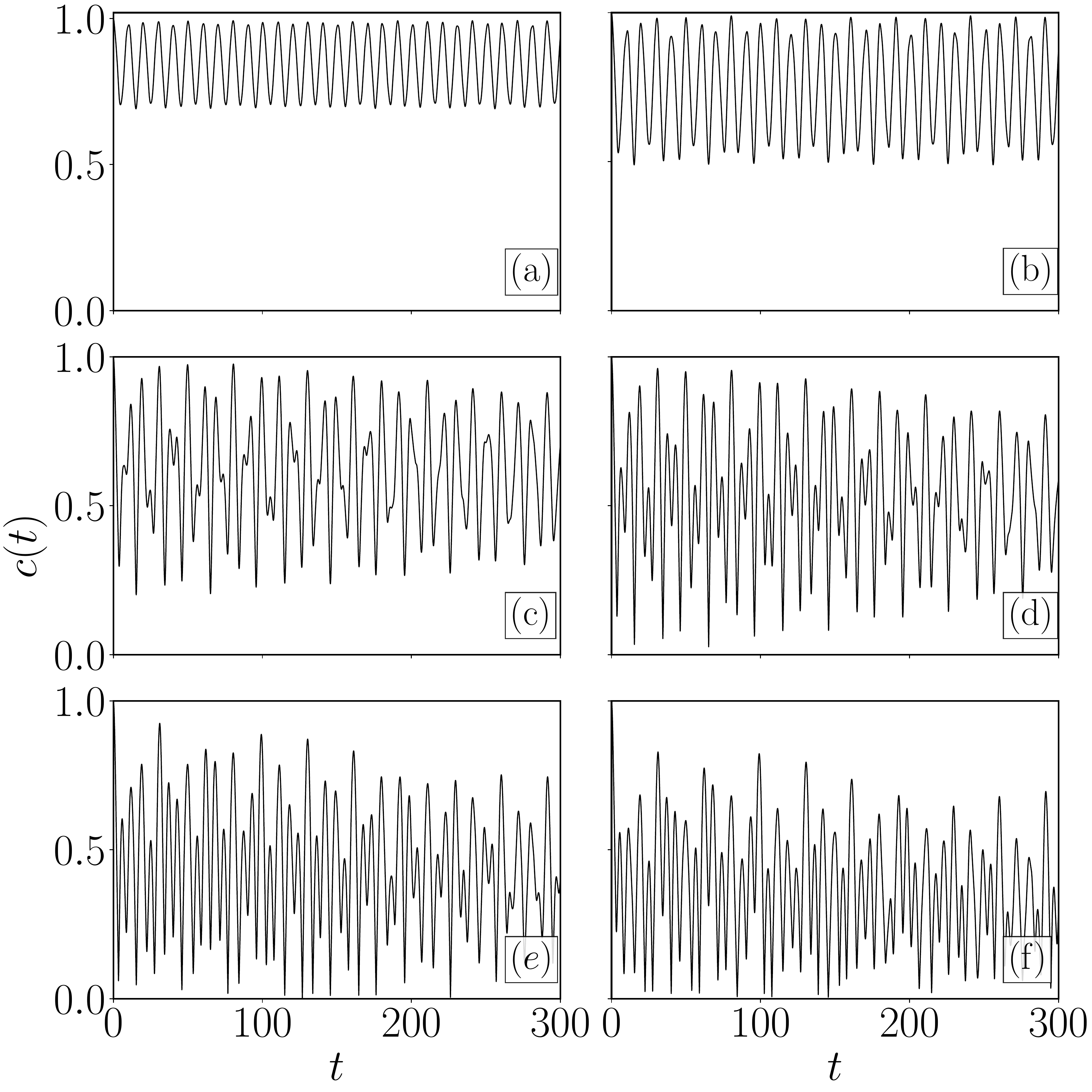}
\caption{Variation of time correlation function with time at $V_0 = 0.5$ after quenching of steady state from (a) $g = 1.4$, (b) $g = 1.6$, (c) $g = 3$, (d) $g = 4$, (e) $g = 6$, and (f) $g = 10$. The other parameters are the same as in Fig.~\ref{fig13}. In the localized state ($g \lesssim 4$), the correlation function exhibit either periodic ($g=1.4$) or quasiperiodic ($g=1.6,2,3$), which becomes chaotic after quenching the system from higher nonlinearity.}    
\label{fig17}
\end{figure}
%%%%%%%%%%%%%%%%%%%%%%%%%%%%%%%%%%%%%%%%%%

We characterize the condensate dynamics by analyzing the temporal evolution of $c(t)$ for different values of $g$, which are plotted in Fig.~\ref{fig15}. Figure~\ref{fig15}(a) illustrates the periodic temporal evolution of $c(t)$ with a period $T \approx 1.309$ for the localized state upon quenching the nonlinearity from $g =1.6 \rightarrow 0$. Figure~\ref{fig15}(b) shows the evolution of $c(t)$ for the localized state when the $g$ is quenched as $g = 2 \rightarrow 0$. The corresponding dynamics show the quasi-periodic oscillation with the presence of two frequencies, which becomes more pronounced for $g=3$ as depicted in Fig.~\ref{fig15}(c). As we analyze the quench dynamics for the state at higher nonlinearity ($g \gtrsim 5$) for which the ground state exhibits a delocalized nature, we find that the corresponding $c(t)$ exhibits aperiodic or chaotic oscillation [see Figs.~\ref{fig15}(d)-(f)]. These dynamical behaviours for different states will become clear as we investigate the PSD of the time correlation, which we discuss below.

%%%%%%%%%%%%%%%%%%%%%%%%%%%%%%%%%%%%%%%%%%%%%%%%%%%%%%%%%%%%%%%%%%%%%%%%%
\begin{figure}[!htb]
\includegraphics[width=1.0\linewidth]{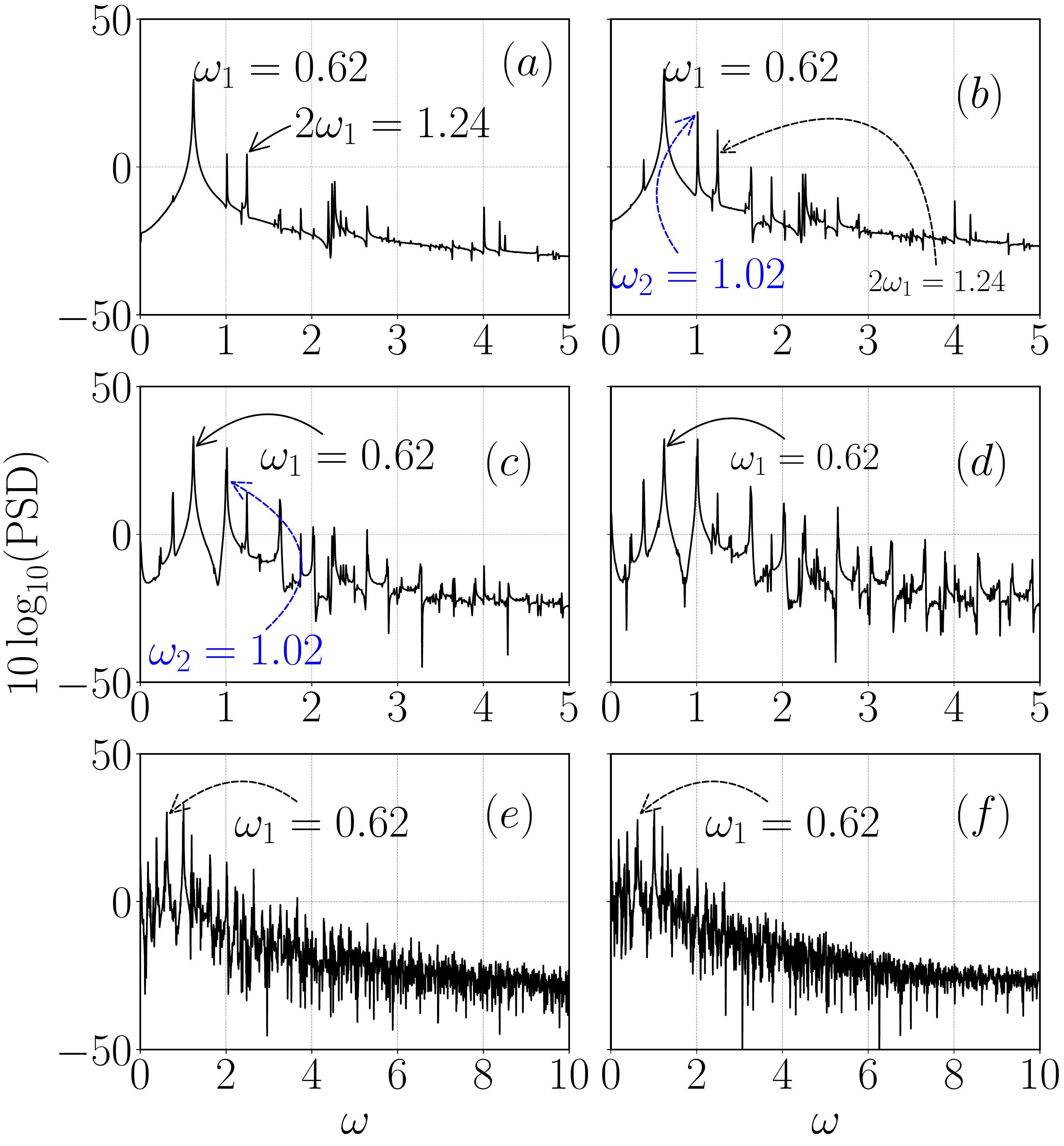}
\caption{The PSD of time correlation function (as shown in Fig.~\ref{fig17}) at $V_0 = 0.5$ after quenching at different nonlinearities, (a) $g = 1.4$, (b) $g = 2$, (c) $g = 3$, (d) $g = 4$, (e) $g = 6$, and (f) $g = 10$ to $g = 0$. Increasing the nonlinearity generates two incommensurate frequencies $\omega_1 = 0.62$, $\omega_2 = 1.02$ at $g \sim 2$. Here, frequencies get populated near $\omega_1$ and $\omega_2$ when we quench the system from low nonlinearity rather than $V_0 = 1.0$.}
    \label{fig18}
\end{figure}
%%%%%%%%%%%%%%%%%%%%%%%%%%%%%%%%%%%%%%%%%%%%%%%%%%%%%%%%%%%%%%%%%%%
In Fig.~\ref{fig16}, we plot the PSD of the time correlation function presented in Fig.~\ref{fig15}. The PSD for the localized state quenched from $g = 1.6\rightarrow0$ is illustrated in Fig.~\ref{fig16}(a). It shows the presence of fundamental frequencies at $\omega_1 =1.13$ along with its higher harmonics like $2\omega_1 = 2.26$, confirming the dynamics to be periodic. However, a quenching from $g = 2 \rightarrow 0$ generates another frequency at $\omega_2=4.09$ along with $\omega_1$, as shown in Fig.~\ref{fig16}(b), indicating the quasiperiodic nature of dynamics of the condensate. The dynamics of the localized states with higher nonlinearity ($g=3$) exhibit the generation of more frequencies around the two frequencies $\omega_1$ and $\omega_2$, as illustrated in \ref{fig16}(c-d). However, for $g\gtrsim 5$, more frequencies start appearing near $\omega_1$ and $\omega_2$ in the PSD. The condensate has a delocalized ground state for $g \gtrsim 5$, shows the presence of a wide range of the frequencies in the dynamics, and the corresponding PSD shows the exponential variation with the angular frequency; a typical signature of the chaotic dynamics. The PSD plots for $g=6$ [Fig.~\ref{fig16} (e)] and $g=10$ [Fig.~\ref{fig16} (f)] indeed show the exponential distribution with the frequencies. Interestingly, similar to the case of quasiperiodic potential, we find that the delocalized state exhibits a chaotic state when the condensate is trapped in random Gaussian disordered potential. The quasiperiodic route to the chaos that exists in the dynamics is similar to those obtained with the condensate trapped in the quasiperiodic potential. 

As we decrease the strength of the random Gaussian disordered potential,  the critical value of the nonlinearity at which the chaotic behaviour appears in the time correlator decreases.  Fig.~\ref{fig17} depicts $c(t)$ at different nonlinearity for the disorder strength $V_0=0.5$ of the \textcolor{blue}{random Gaussian disordered potential}. In this case, the nature of $c(t)$ is periodic for $g=1.4$, quasiperiodic for $g=2,3$, and chaotic for $g=4$ and $6, 10$. As we analyze the corresponding PSD, we find the presence of fundamental frequency at $\omega_1=0.62$ (as shown in Fig.~\ref{fig18}(a)), which is lower than those for $V_0=1$ which is $\omega_1 = 1.12$. An increase in $g=2$ leads to the generation of other frequency $\omega_2=1.02$, which is incommensurate with the fundamental frequency ($\omega_1$),  indicating the quasiperiodic nature of the dynamics [see Fig.~\ref{fig18}(b)]. Further, an increase in the nonlinear interaction to $g=3$ [Fig.~\ref{fig18}(c)], and $g=4$ [Fig.~\ref{fig18}(d)] generates other frequencies, which origin can be understood as a combination of $\omega_1$ and $\omega_2$. However, for $g=6$ and $g=10$, the PSD exhibits exponential distribution with the frequencies indicating the fully chaotic state. Interestingly, the route to chaos for $V_0=0.5$ remains a quasiperiodic route similar to what we obtained for $V_0=1$. 
%%%%%%%%%%%%%%%%%%%%%%%%%%%%%%%%%%%%%%%%%%%%%%%%%%%%

\begin{figure}[!t]
\includegraphics[width=0.85\linewidth]{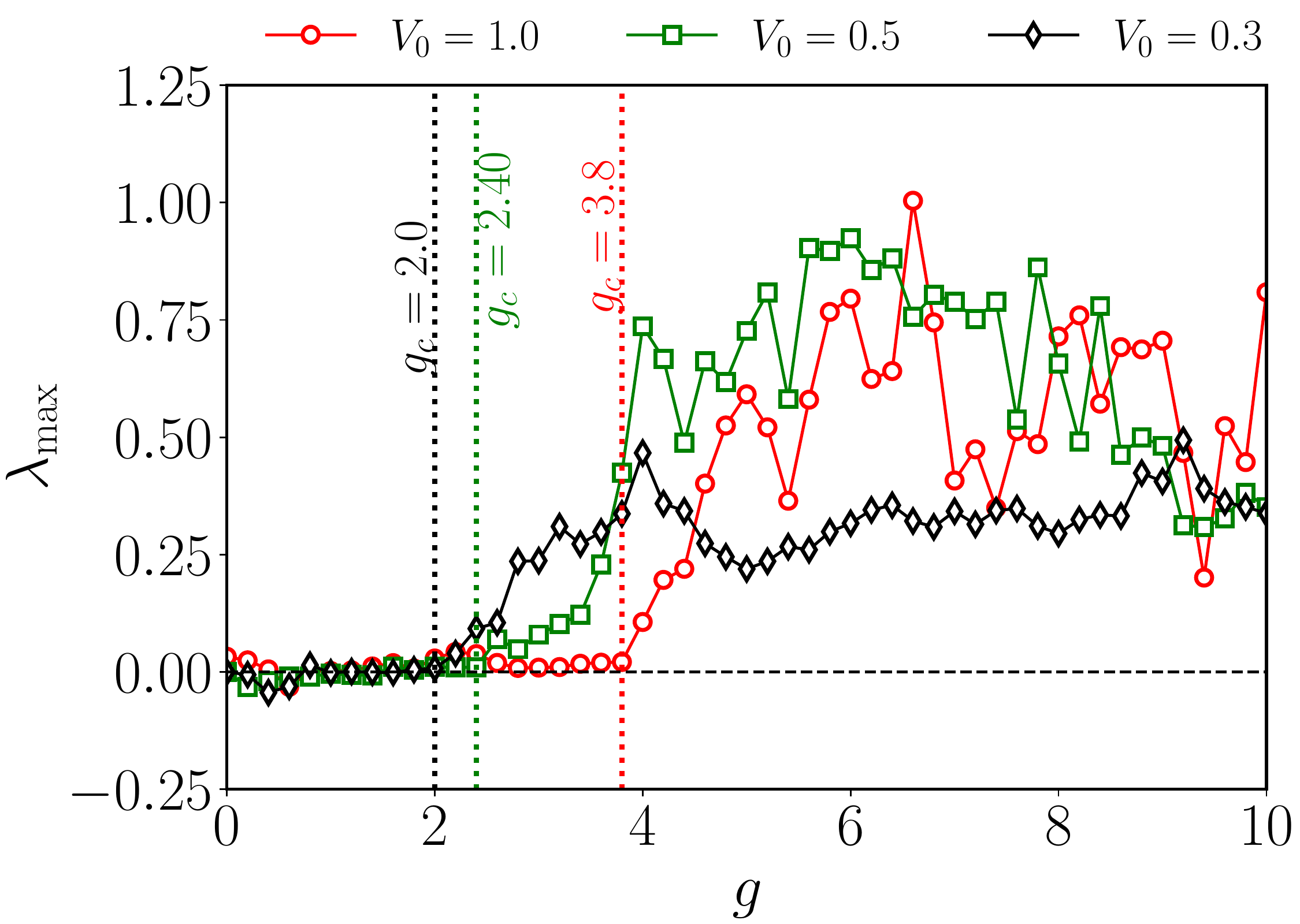}
\caption{Plot showing the maximal Lyapunov exponent ($\rm \lambda_{max}$) averaged over five different random realizations, plotted against nonlinearity for different strengths of the random potential: $V_0=1$ (red), $V_0=0.5$ (green), and $V_0=0.3$ (black). The localized state has $\rm \lambda_{max} \lesssim 0$, while the delocalized states have $\rm \lambda_{max}>0$, indicating chaotic dynamics. The threshold $g_c$ above which $\rm \lambda_{max}>0$, a characteristic of the delocalized state, decreases as $V_0$ decreases. Vertical dotted lines are included to guide the eyes for different $g_c$ values. The $g_c$ value is evaluated by taking the average over five realizations. }
\label{fig19}
\end{figure}
%%%%%%%%%%%%%%%%%%%%%%%%%%%%%%%%%%%%%%%%%%%%%%%%%%%%%%%%%%%%%
%\textcolor{blue}{After associating the delocalized state with the chaotic dynamics, now, we focus on complementing the studies by computing the maximal Lyapunov exponent ($\rm \lambda_{max}$) of the time correlator function $c(t)$. In case of random Gaussian disordered potential we considered single disorder realization to show the density and time correlator function. To check the degree of universality,  we compute the maxmimal Lyapunov exponent by taking the average of five different disorder realizations of $c(t)$ for each set of $V_0$. For a given $V_0$ in Fig.~\ref{fig19} the average $\rm \lambda_{max}$ remains negative for the delocalized state. For instance, for $V_0=1$, the $\rm \lambda_{max} \sim 0$ for $g \lesssim 3.8$. However, beyond this nonlinearity, $\rm \lambda_{max}$ becomes positive and remains above zero for a higher value of $g$. The threshold value of the nonlinearity for $V_0=1$ indicates the delocalized feature of the condensate. Decreasing the potential strength to $V_0=0.5$ results in the decrease of the threshold nonlinearity to $g_c\sim 2.40$. Further, decrease to $V_0=0.3$ makes the $g_c \sim 2.0$.}
%%%%%%%%%%%%%%%%%%
After associating the delocalized state with the chaotic dynamics, we now focus on complementing the studies by computing the maximal Lyapunov exponent ($\rm \lambda_{max}$) of the time series $c(t)$. In Fig.~\ref{fig19}, we plot the variation of $\rm \lambda_{max}$ \textcolor{blue}{averaged over five different random realizations} with $g$ for different sets of $V_0$. For a given $V_0$, $\rm \lambda_{max}$ remains negative or close to zero for the localized state, while it becomes positive for the delocalized state. For instance, for $V_0=1$, the $\rm \lambda_{max} \sim 0$ for $g \lesssim 3.8$. However, beyond this nonlinearity, $\rm \lambda_{max}$ becomes positive and remains above zero for a higher value of $g$. The threshold value of the nonlinearity for $V_0=1$ indicates the delocalized feature of the condensate. Decreasing the potential strength to $V_0=0.5$ decreases the threshold nonlinearity to $g_c\sim 2.4$. Further, decrease to $V_0=0.3$ makes the $g_c \sim 2.0$.
%\textcolor{green}{Note that, the $\rm \lambda_{max}$ is averaged over five different disorder realizations and the analysis showed before is obtained using a single disorder realization. Therefore, the $g_c$ value at which the dynamics become chaotic it might get different from the approximated value estimated from Fig.~\ref{fig16} and Fig.~\ref{fig18}.}

\begin{figure}
    \centering
    \includegraphics[width=0.85\linewidth]{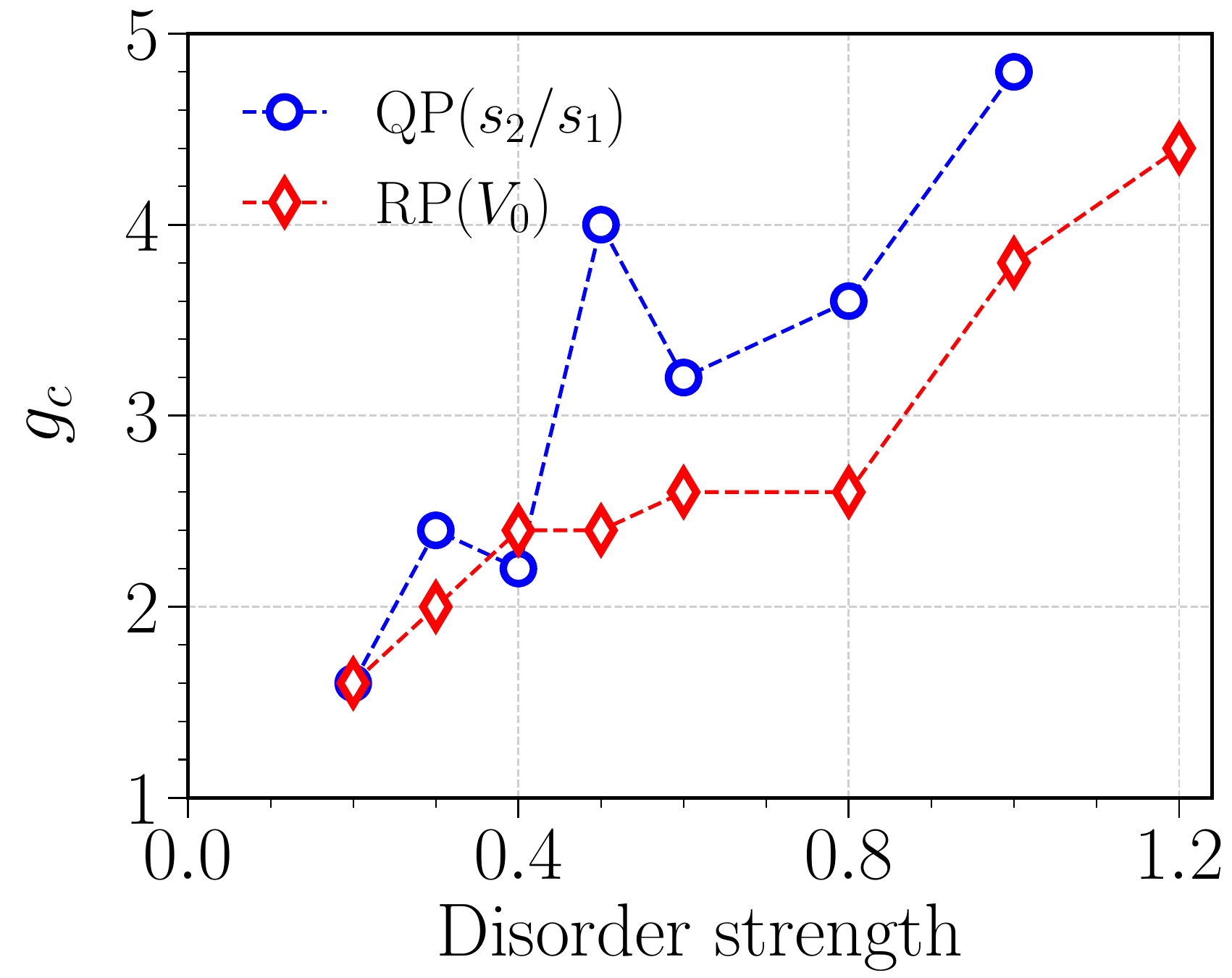}
    \caption{Plot showing the variation of the critical non-linearity strength ($g_c$) with the disorder strength ratio $s_2/s_1$ for the quasi-periodic (QP) lattice and the potential strength $V_0$ for the random potential (RP).}
    \label{fig:gc}
\end{figure}

Next, in Fig.~\ref{fig:gc}, we present a comparative analysis of the $g_c$ at which the dynamics of the condensate begin exhibiting a positive Lyapunov exponent for both the quasiperiodic optical lattice (blue circles) and the random Gaussian disordered potential (red diamonds). We observe that the critical value $g_c$ increases for both potentials as the disorder strength increases. Although the $g_c$ appears to be of the same order for both disorder potentials at low disorder strengths, for higher disorder strengths, $g_c$ is higher for the quasi-periodic potentials than for the random Gaussian potentials.

%%%%%%%%------------------------
%\begin{figure}
 %   \centering
 %   \includegraphics[width=\linewidth]{log-dis-fun-freq-BOL.pdf}
 %   \caption{Variation of fundamental frequency $\omega_1$ with $s_2/s_1$ for bichromatic optical lattice. (b) Variation of fundamental frequency $\omega_1$ with $s_2/s_1$ for random speckle potential. Fundamental frequency obtained from the Power spectral density of respective time correlation function. The red dashed line associated to the quadratic polynomial and black solid line corresponds to linear fit of the respective frequencies. }
  %  \label{fig:my_label}
%\end{figure}
%%%%---------fit curve
%\begin{figure}
%    \centering
%    \includegraphics[width=\linewidth]{log-dis-fun-freq-RSL.pdf}
%    \caption{Variation of fundamental frequency $\omega_1$ with $V_0$ for random speckle potential in logarithmic scale. Fundamental frequency obtained from the Power spectral density of respective time correlation function. The frequencies follows power law behaviour with disorder strength }
 %   \label{fig:my_label}
%\end{figure}
%% The Table~\ref{table1} illustrates proportional relation between quenched nonlinearity and disordered strength.    

\section{Summary and Conclusion}
\label{sec5}
In this paper, we have studied the effect of atomic interaction on the ground state and the associated dynamics of Bose-Einstein condensates in a one-dimensional bichromatic optical lattice and random Gaussian disordered potentials. We identified that increasing the nonlinearity strength leads to the delocalization of the condensates. We have analyzed the condensate dynamics by quenching the nonlinearities to zero from the value at which we prepare the ground state. We noticed regular dynamics of the condensate for small nonlinear strengths, while it becomes chaotic at large nonlinearities where delocalization occurs. We also identified a quasiperiodic route to chaos for both bichromatic and random Gaussian disordered potential. The power spectral density displays a broadband spectrum, and the maximal Lyapunov exponent is positive when it exhibits chaotic dynamics. The power spectral density and largest Lyapunov exponent confirm the presence of chaotic dynamics. Further, we have found that the critical nonlinearity for delocalization decreases by decreasing the ratio of amplitudes of the secondary to primary laser for quasiperiodic potential or the strength of the random Gaussian disordered potential. Our studies upon quenching the nonlinear interaction reveal regular dynamics of the condensate for the localized state while it becomes chaotic for the delocalized state. In this study, we have restricted our analysis to the scalar BECs. However, it would be interesting to extend the work for the spin-orbit coupled spinor BECs where the quenching of coupling parameters brings a similar effect as quenching of the nonlinear and thus have the possibility of richer dynamics~\cite{Gangwar2022}. Also, it would be interesting to extend the work for the finite quenching rate.

\section{Acknowledgments}
\label{sec6}
We thank Kanhaiya Pandey, Sadhan K. Adhikari, Luca Salasnich, and Saptarishi Chaudhuri for the fruitful discussions and suggestions. We also gratefully acknowledge our supercomputing facility Param-Ishan (IITG), where all the simulation runs were performed. The work of PM is supported by DST-SERB under Grant No. CRG/2019/004059, DST-FIST under Grant No. SR/FST/PSI-204/2015(C), and MoE RUSA 2.0 (Physical Sciences). PKM acknowledges the Department of Science and Technology - Science and Engineering Research Board (DST-SERB) India for the financial support through  Project No. ECR/2017/002639.

\bibliography{references_chaos.bib}
\end{document}